\documentclass{article}
\usepackage{pslatex}
\usepackage{graphics}
\usepackage{rotating}
\usepackage[table]{xcolor}
\usepackage{caption}
\usepackage{hyperref}
\usepackage{amsthm}
\usepackage{enumitem}
\usepackage{empheq}
\theoremstyle{plain}

\begin{document}

\title{The Less Intelligent the Elements, the More Intelligent the Whole. Or, Possibly Not?}

\author{Guido Fioretti\\University of Bologna}
\maketitle

\begin{abstract}
From neural networks to the immune system to agent-based models, connectionism aims at deriving complex behavior from relatively simple components. However, one unsettled question is how ``intelligent'' these components should be, and in what ways their local intelligence relates to the emergence of collective intelligence.

I approach this problem by endowing the preys and predators of the Lotka-Volterra model with behavioral algorithms characterized by different levels of sophistication, identified by extensive exploration of the relation between individual and collective intelligence across numerous disciplines. The main finding is that by endowing both preys and predators with the capability of making predictions based on linear extrapolations that exploit their knowledge of Lotka-Volterra dynamics, a novel sort of dynamic equilibrium appears, where both species coexist while both populations grow indefinitely. This explosive dynamics is coherent with economic interpretations of the Lotka-Volterra model.

While this finding does not invalidate the connectionist philosophy that relatively simple components can generate complex outcomes, it also suggests that certain interesting macroscopic outcomes can only be generated by components that are not so simple, after all.
\end{abstract}


\newpage

\section{Introduction}     \label{sec:intro}

Connectionism has revolutionized cognitive sciences with the idea that intelligent macroscopic behavior can arise out of the structure of connections between relatively simple components, as brain neurons are \cite{smolensky-88BBS} \cite{clark-93}. By analogy, it has inspired similar approaches in other domains where interactions between a large number of elements generate qualitatively more complex macroscopic behavior, such as the immune system \cite{farmer-90PD} \cite{nunesdecastro-vonzuben-dedeus-03N} or human societies \cite{read-miller-98tutto} \cite{vanoverwalle-11XVI}. In particular, since human decision-makers can often perform computations that are more complex than weighted summations as artificial neurons do, 
Agent-Based Models (ABMs) have been employed in the social domain, rather than neural networks \cite{fioretti-16II}.

In general, ABMs have been employed as sufficiency proofs that would identify the minimalist microscopic conditions for  specific macroscopic dynamics to appear \cite{fioretti-12ORM}, a posture succinctly described by the KISS principle (Keep It Simple, Stupid)  \cite{epstein-axtell-96}. This  clearly derives from the connectionist philosophy of obtaining complex behavior out of interactions between relatively simple elements, to be applied this time to interacting (simulated) humans instead of interacting (simulated) neurons \cite{weick-roberts-93ASQ} \cite{clark-chalmers-98A} \cite{trianni-tuci-passino-marshall-11SI}. 

However, growing diffusion and  realism of ABMs has eventually suggested that,  in order to obtain meaningful results, artificial agents should not be too simple. The alternative KIDS principle  (Keep It Detailed, Stupid) proposes that artificial agents should be endowed with fairly sophisticated capabilities in order  to generate relevant macroscopic regularities  \cite{edmonds-moss-05XI}. Thus the question arises, how ``intelligent'' should artificial agents be. Do criteria exist, that allow  modelers to estimate how detailed their artificial agents should  be, in order to generate meaningful emergent properties? Some attempts have been made at resolving this debate by comparing several ABMs, but results have been largely inconclusive hitherto \cite{edmonds-08VIII} \cite{sun-lorscheid-millington-lauf-magliocca-groeneveld-balbi-nolzen-muller-schulze-buchmann-16EMS}.

I approached this conundrum from a slightly different point of view. I rather focused on one single model, namely an ABM version of the Lotka-Volterra prey-predator model (LV) \cite{lotka-25} \cite{volterra-26N}. One reason to select this model is that the LV applies to many social dynamics as well; for instance, just like in the LV predators feed on preys but starve once preys have (almost) disappeared, salaries draw from capital gains but fall if those become exhausted, or  ideologies fade if their opponents disappear \cite{choi-97NDPLS} \cite{wijeratne-yi-wei-09CSF} \cite{wu-liu-wang-12TFSC} \cite{christodoulakis-16DPE} \cite{marasco-romano-18QQ} \cite{hidayati-kurniawan-21IJERSS} \cite{wang-chen-wu-21DDNS}. Within the framework of this specific model I explored specific forms of individual intelligence that have been suggested by empirical findings, theoretical statements and numerical simulations  across disciplines as being capable of generating some form of collective intelligence.

The rest of this article is organized as follows. In the ensuing section \S~\ref{sec:individual-collective}  I  distill propositions  from several disciplines concerning the relations between individual and collective intelligence. In \S~\ref{sec:prey-predator} I translate these propositions into behavioral algorithms that can be ascribed to either predators, or preys, or both. Unexpectedly, besides behaviors that generate co-existence of stable populations there exists one behavioral algorithm that generates co-existence of exploding populations of both predators and preys, a circumstance that I discuss in the concluding section \S~\ref{sec:conclusions} as remindful of unlimited growth in capitalistic economies. The mathematical properties of the LV model, the  parameters and outputs of our model and a sensitivity analysis of its results are expounded in appendices \S~\ref{app:lyapunov},  \S~\ref{app:agent-based-LV} and  \S~\ref{app:sensitivity}, respectively.~\footnote{The  \href{www.comses.net/codebases/0eada5b3-3d18-4fc6-92af-841ff0971d28/}{code}  is available  on \href{https://www.comses.net/}{CoMSES}. This code has been written by Andrea Policarpi upon inspiration from Wilensky and Reisman's \emph{Wolf Sheep Predation} model \cite{wilensky-97} \cite{wilensky-reisman-06CI}, available among the library models of  the \href{https://ccl.northwestern.edu/netlogo}{NetLogo} platform.}

\section{Individual and Collective Intelligence}  \label{sec:individual-collective}

In this section I review theoretical insights and empirical evidence that bear on the question whether a collectivity works better if their members understand its underlying mechanisms, or rather enact simple rules. This survey aims at being as broad as possible, ranging from the computational to the social sciences and beyond, including the neurosciences as well as group psychology and  psychoanalysis. Its findings are essential in order to carry out a reasoned, rather than random exploration of behavioral algorithms for predators and preys.

I condensed the results of this survey in three propositions, the third of which comes in three versions. Henceforth I list these propositions  in ascending order, from purely KISS- to increasingly KIDS-supportive. Subsequently, these propositions are complemented by three technical qualifications   in \S~\ref{subsec:qualifications}.

The first proposition subsumes positions that justify the KISS principle on theoretical and empirical grounds. To varying degrees, these positions imply that KISS is not just a convenient  assumption but also a reliable guideline to describe what humans actually do (descriptive value), or at least what they should do (normative value).

One extreme version  maintains that individual intelligence is totally irrelevant for social sciences \cite{ormerod-08V}. Specifically, several economic ``zero-intelligence'' models make this assumption to describe double-auction  markets \cite{gode-sunder-93JPE} \cite{gode-sunder-97QJE},  financial markets \cite{farmer-patelli-zovko-05PNAS} and, more in general, any sort of prediction-based market \cite{othman-08CVI}. Such models  typically  suggest that markets --- even futures markets --- can work perfectly well with decision makers who have no memory and are incapable of  whatsoever form of  cognition. However, some of these models have been found to conceal constraints that artificially generate their most impressive results \cite{gjerstad-shachat-21NDPLS}.

A less extreme version identifies \emph{Fast and Frugal Heuristics} (FFH) that are reasonably successful, robust, computationally cheap and widely employed in the practice of decision-making \cite{gigerenzer-todd-99} \cite{raab-gigerenzer-15FP}\cite{mousavi-schulkin-19XIII} \cite{forbes-igboekwu-mousavi-20} \cite{flyvbjerg-24PMJ}. This research stream stresses that simple heuristics are in general more  effective than complex optimization procedures.

Notably, the FFH paradigm does not understand bounded rationality \cite{simon-57} \cite{simon-82} as a deviation from optimality but rather as the evolutionary optimal answer  to radical uncertainty and imperfect information \cite{rieskamp-hertwig-todd-06XI} \cite{hertwig-herzog-09SC} \cite{gigerenzer-16I} \cite{mousavi-gigerenzer-17HO}. Hence, it implies that heuristics have a descriptive as well as a prescriptive value \cite{gigerenzer-02XXII} \cite{hutchinson-gigerenzer-05BP} \cite{wade-hands-14JEM} \cite{descioli-kurzban-todd-15IX}.

Similar conclusions are suggested by  \emph{The Traveler's Dilemma}, a game where rational players endowed with unlimited ability to read one other's minds (i.e., ``I think that she thinks that I think that ...'') receive lower payoffs than boundedly rational players who carry out mind reading for a couple of steps at most \cite{basu-94AER} \cite{basu-07SA}. Interestingly, when this game is played with real people they invariably behave the ``stupid,'' more efficient way.

Finally, ABMs of the stock market where  agents with diverse degrees of ``intelligence'' operate have generated the unexpected finding that sophisticated agents typically make lower profits than the simpler ones   \cite{terna-02VIII} \cite{gou-06JASSS}. The reason was that  sophisticated estimation techniques suffered from over-fitting  a  noisy environment; by contrast, simple rules traded-off  imperfections with  robustness.

The following proposition summarizes  KISS-supportive stances in a  strong, prescriptive, though not universal claim:

\newtheorem{theorem}{Proposition}
\begin{theorem}\ \\  \label{proposition:betterStupid}
There exist settings where lack of individual intelligence is necessary  to reach collective intelligence.
\end{theorem}

However, other statements point in the opposite direction. Specifically, the ability to make predictions surfaced quite often as an aspect of individual intelligence that has a substantial impact on collective intelligence, where prediction can be meant to be very  local and very limited, such as linear extrapolation,  or  it may involve extremely sophisticated cognitive abilities. Correspondingly, such statements can span the whole gamut from moderate KISS to maximum KIDS.

Among simulations, the ability to predict global behavior generally enables agents to improve their performance  \cite{kephart-hogg-huberman-90PD} \cite{kephart-huberman-hogg-92IV} \cite{scheutz-schermerhorn-03LIV}. However, macro-to-micro backward causation may eventually curb this effect \cite{garciadiaz-24NDPLS}.

Shifting from simulations to the real world,   human groups  act as teams insofar their members make the effort of reading one  other's mind, predict what their mates will do, and behave accordingly \cite{klein-98} \cite{walker-18}. This is confirmed by experimental tests  on indicators of social attentiveness, such as eye movements or taking turns at speaking \cite{williamswooley-chabris-pentland-hashmi-malone-10S}.

Such experiences suggest that the individual ability to make predictions can contribute to collective intelligence. By expressing a potentiality, the following proposition can accomodate several stances in the KISS--KIDS gamut:

\begin{theorem}\ \\  \label{proposition:prediction}
Individual intelligence, if employed in order to predict and coordinate  behavior, can improve collective intelligence.
\end{theorem}

Finally, a third stream  points to the ability to think and understand the dynamics of collective behavior as the  maximum sophistication that  individuals can attain \cite[p.~292]{bateson-58}  \cite[p.~21]{argyris-schon-96} \cite{visser-04WP}. In psychology and elsewhere, this ability is known as  \emph{meta-cognition} \cite{klein-98}.

Meta-cognition represents maximum KIDS.  Its common core can be expressed as follows:

\begin{theorem}\ \\  \label{proposition:meta-cognition}
Single individuals may be able to envisage  the collective intelligence of their organization.
\end{theorem}

Notably, given an individual capable of meta-cognition, their possibilities for affecting the organization depend  on  hierarchical position \cite{ortenblad-koris-13IJEM} \cite{li-yang-shi-21H}. Thus,   Proposition~\ref{proposition:meta-cognition} can be found in three versions that differ from one another in this respect.

The first version of Proposition\S~\ref{proposition:meta-cognition} is  \emph{double-loop learning} \cite{argyris-schon-78}. Also known   as ``higher-level'' \cite{fiol-lyles-85AMR}, ``adaptive'' \cite{senge-90}, ``radical'' \cite{miner-mezias-96OS}, ``second-order'' \cite{arthur-aymansmith-01AMJ} or ``meta'' \cite{argyris-03OS} -learning, it has the action-oriented flavor of management studies where one or a few leaders understand causal relations and conceive innovative solutions. This point of view can be expressed as follows:

\newtheorem{corollary}{Proposition}[theorem]
\begin{corollary}  \label{proposition:meta-cognition_double-loop}
Single individuals may be able to envisage  the collective intelligence of their organization and steer it towards  alternative dynamics.
\end{corollary}

By contrast, if those individuals who envisage the collective intelligence of their organization do not have the authority to steer it,  \emph{deutero-learning} ensues \cite{bateson-58} \cite{bateson-72}. Unsurprisingly, this concept stems from anthropology rather than management. It  rather fits the case of individuals who must passively accept a mechanism of which they become aware, but whose aims they do not share \cite{visser-03JHBS} \cite{visser-07AMR}. This point of view can be expressed by the following version of Proposition~\ref{proposition:meta-cognition}:

\begin{corollary}  \label{proposition:meta-cognition_deutero}
Single individuals may be able to envisage  the collective intelligence of their organization while not being able to steer it towards  alternative dynamics.
\end{corollary}

Finally, a still different case occurs when the collective intelligence of an organization is envisaged not by one or a few individuals who may or may not be able to change it, but rather by the vast majority of its individuals who openly agree on a new course of action \cite{kohn-99} \cite{wong-wei-yang-tjosvold-17JWB}. This third version of Proposition~\ref{proposition:meta-cognition} can be expressed as follows:

\begin{corollary}   \label{proposition:meta-cognition_agreement}
The majority of individuals may be able to envisage  the collective intelligence of their organization and agree to steer it towards alternative dynamics.
\end{corollary}

\subsection{Technical Qualifications}     \label{subsec:qualifications}

The  emergence of collective dynamics out of individual behavior has also technical features that are unrelated to the degree of individual intelligence. These technical features can  eventually qualify the above propositions.

The first qualification requires individuals to be heterogeneous in order to generate interesting collective dynamics. This qualification applies even to extreme KISS approaches such as artificial neural networks, whose neurons are as simple as summators and yet require random initialization of their weights. In more complex models, heterogeneity increases the range of possible computations  \cite{saffre-hildmann-deneubourg-18SI} \cite{sanchezpuig-zapata-pineda-iniguez-gershenson-23FCS}- One example is provided by the electrical signals generated by biological neurons, which fire in unison  under epileptic seizures  but exhibit complex patterns otherwise \cite{beenhakker-huguenard-09N} \cite{herrmann-demiralp-05CN}. 

Moving on to collective behavior, group psychoanalysis  identifies three irrational collective behavior modes characterized by individuals merging their minds into a homogeneous, not-so-intelligent whole as, e.g., in  religious or political sects \cite{bion-61} \cite{tchelebi-17OSD}. By contrast, crowd wisdom --- the ability of large groups to provide solutions that would be difficult for isolated individuals to reach \cite{johnson-00ANYAS} --- relies on  heterogeneity of individual interpretations  \cite{hong-page-09JET} \cite{hong-page-12III}.

The above considerations can be subsumed by the following, rather simple qualification:

\newtheorem{theorem1}{Qualification}
\begin{theorem1}   \label{qualification:heterogeneity}
Individual heterogeneity  is necessary  to reach collective intelligence.
\end{theorem1}

The second  qualification concerns collective agreements. Thus, it is  particularly relevant for Proposition \S~\ref{proposition:meta-cognition_agreement}.

This qualification makes the point that although collective agreements necessarily generate stable equilibria,  individual intelligence can still make a difference insofar it concerns the speed of convergence.

Macroeconomics has a good point in case. In the 1970s,  many Central Banks were trying to stimulate the economy by increasing the money supply  in the belief that economic actors would use it to increase GDP. However,  those actors were eventually able to anticipate that inflation would be the long-term consequence, which expectation contributed to generate inflation  \cite{lucas-76}. In this case, greater individual intelligence --- known in economics  as \emph{rational expectations} --- made for faster convergence to a high-inflation, high-unemployment equilibrium (note that in this case the ``agreement'' exists only among  actors other than the Central Bank, which is treated as  exogenous).

This insight can be expressed by the following qualification:

\begin{theorem1}   \label{qualification:speed}
Given a stable equilibrium generated by a collective agreement,  individual intelligence accelerates convergence.
\end{theorem1}

Finally, the third technical qualification concerns the possibility that small individual decisions have a sizable impact on collectives. Specifically, the \emph{butterfly effect} states that non-linearities, through chains of bifurcations, are capable of generating  chaotic dynamics \cite{lorenz-72AAAS}. 
This  qualification can be expressed as follows:

\begin{theorem1}   \label{qualification:butterfly}
Any degree of individual intelligence can trigger non-linear interactions that ultimately have a substantial impact on collective intelligence.
\end{theorem1}

\section{What if Preys and Predators Think?}  \label{sec:prey-predator}

This section   endows the predators and preys of the LV model  \cite{lotka-25} \cite{volterra-26N} with variants of individual intelligence inspired by the propositions and qualifications outlined in \S~\ref{sec:individual-collective}. In particular, this exploration aims at identifying behavioral algorithms that allow co-existence of predators and preys.

The LV model, also known as the prey-predator model, is based on two animal species that  take up competing roles but ultimately need one another. On the one hand,  preys would grow unchecked if predators would not exist. On the other hand, predators would starve if they would succeed to eliminate all the preys. Thus, in the LV  co-existence of predators and preys is a systemically desirable state.

Species co-existence in the LV is an attractive instance of collective intelligence for at least two reasons. The first one is that this model is relevant to a number of settings other than animal species, including as diverse applications as the business cycle \cite{goodwin-67IV} \cite{malcai-biham-richmond-solomon-02PRE} \cite{solomon-richmond-02EPJB} \cite{wu-liu-wang-12TFSC}, technological substitution \cite{bhargava-89TFSC} \cite{morris-pratt-03TFSC} \cite{wu-liu-wang-12TFSC}, ideological and power struggles \cite{vitanov-dimitrova-ausloos-10PA} \cite{christodoulakis-16DPE} \cite{marasco-romano-18QQ} \cite{hidayati-kurniawan-21IJERSS} and market-share dynamics \cite{wijeratne-yi-wei-09CSF} \cite{marasco-picucci-romano-16TFSC} \cite{wang-chen-wu-21DDNS}.

Consider, for instance, the interpretation of business cycle dynamics in terms of the LV model \cite{goodwin-67IV}. Suppose that income is shared between capital gains that are entirely re-invested, and salaries that are entirely consumed. Investments make the economy grow, but the closer the economy comes to full employment, the easier it is for salaried labor (the ``predators'') to squeeze profits until companies (the ``preys'') go bankrupt --- or relocate --- until unemployment increases again, enabling the economy to start growing again \cite{marx-67}.

Similar dynamics can represent a variety of social relations characterized by some form of antagonism between actors that, for one reason or another ultimately need one another. Exploring the possibility that predators and preys are capable of sophisticated forms of individual intelligence  reflects these sorts of applications, rather than the purely biological ones.

The second reason is that although the basic LV has a family of equilibria where species coexist through  cycles, these equilibria are structurally unstable. Very simple additions to the basic model such as random noise \cite{mao-sabanis-renshaw-03JMAA} \cite{zhu-yin-09JMAA} \cite{liu-fan-17JNS} \cite{vadillo-19AMC}, time delays \cite{zhen-ma-02NA} \cite{bahar-mao-04JMAA} \cite{yan-chu-06JCAM} or discrete dynamics \cite{reichenbach-mobilia-frey-06PRE} \cite{parker-kamenev-09PRE} are sufficient to drive the LV away from the equilibrium cycle.

In particular, ABM versions of the LV model reflect the finding that the above additions destroy the possibility of species co-existence  \cite{reichenbach-mobilia-frey-06PRE} \cite{parker-kamenev-09PRE}. Several ABMs resume co-existence of species by adding \emph{ad hoc} features  such as availability of food and shelter for preys, or  stages of phenotypic development for predators \cite{wilson-96TPB} \cite{wilson-98AN} \cite{wilensky-reisman-06CI} \cite{carmichael-hadzikadic-13ACS} \cite{thierry-sheeren-marilleau-corson-almaric-monteil-15EM}.

By contrast, I pursue the goal of obtaining the collectively intelligent co-existence of predators and preys from some form of intelligence of their individual decisions. On an ABM version of the LV that does not admit co-existence otherwise.

Let us first consider the basic LV model. Let~$x \in \Re^+_0$ and~$y \in \Re^+_0$ denote the environmental densities of preys and predators, respectively~\footnote{\emph{Environmental density} is the ratio of the number of individuals to some measure of the size of the natural environment where  predators and preys live, such as its area or volume.} of  preys and predators, respectively. The LV model captures their dynamics by means of the following pair of differential equations:

\begin{equation}        \label{eq:lotka-volterra}
  \left\{
  \begin{array}{rcl}
    \dot{x} & = & a x - b x y \\
    \dot{y} & = & -c y + d x y
  \end{array}
  \right.
\end{equation}
where $a, b, c , d \in \Re^{+}$ are suitable constants.

This model has three outcomes:

\begin{description}
\item[$E_1 = (0, 0)$] No preys, no predators. If $x(0)=0$ and $y(0)=0$, then $x(t) = y(t) =0 \forall t>0$. Note that this equilibrium is  also approached  if $x(0)=0$ and $y(0)>0$, in which case $x(t)=0 \forall t>0$ and $y(t) = y(0) \, e^{-ct}$, which implies that $\lim_{t \to + \infty} y(t) = 0$. 

\item[$E_2 = (+ \infty, 0)$] Only preys, no predators. If $x(0) >0$ and $y(0)=0$, $x(t) = x(0) \, e^{at}$ and $y(t) = 0 \forall t>0$. The population of preys grows indefinitely because   $\lim_{t \to + \infty} x(t) = + \infty$. The ABM-LV can generate equilibria where only preys are left, that do not grow indefinitely; henceforth, such equilibria will be assimilated to $E_2$.

\item[$E_3 = (c/d, a/b)$] Preys and predators coexist. This equilibrium is globally stable for sufficiently small~$a$, sufficiently large~$b$, sufficiently large~$c$ and sufficiently small~$d$ (see Appendix~\ref{app:lyapunov} for details).

\end{description}

It can be shown that $E_3$ appears if coefficients~$a$ and~$d$ are sufficiently small and coefficients~$b$ and~$c$ are sufficiently large. In other words, preys and predators are more likely to coexist if preys do not grow too quickly and, furthermore, their growth is effectively checked by predators. Predators grow with preys, but  co-existence is achieved because predators   slow down their growth when their own population grows.

Let us turn to the discrete-time, agent-based version  in order to analyze what behavioral algorithms lead the system into either~$E_1$, $E_2$, or~$E_3$, respectively. In particular, I  modified the behavioral algorithms of Wilensky and Reisman's \emph{Wolf Sheep Predation} model \cite{wilensky-97} \cite{wilensky-reisman-06CI}~\footnote{Wilensky and Reisman's model is available in \href{https://ccl.northwestern.edu/netlogo}{NetLogo}, models' library. My  \href{www.comses.net/codebases/0eada5b3-3d18-4fc6-92af-841ff0971d28/}{extended version} is available at \href{https://www.comses.net/}{CoMSES}.}. Parameters have been kept at their original default values (see Appendix~\ref{app:agent-based-LV} for details). Technical Qualification~\ref{qualification:heterogeneity} applies.

In the basic version of Wilensky and Reisman's ABM-LV, $E_3$ almost never sustains itself. Specifically,  $E_1$ (both preys and predators disappear) is observed $41.9\%$ of the times, $E_2$ (only preys survive) $57.9\%$ of the times whereas only $0.2\%$ of the times the two species manage to coexist. While Wilensky and Reisman solved this problem by adding a resource consumed by the preys, I rather endowed preys and predators with  behavioral algorithms inspired by the Propositions and Qualifications expounded in \S~\ref{sec:individual-collective}. 

The ensuing subsections report my explorations. All reported results have been averaged over 1,000 runs (details   in Appendix~\ref{app:agent-based-LV}).

\subsection{It's Difficult to Leave a KISS}   \label{subsec:failingEscape}

Suppose that predators and preys  devise a reproductive strategy which, in their intentions, should be able to keep the system at $E_3$. With a slight departure from pure KISS, they may imagine that a flexible and adaptive strategy  based on some negative feedback   may  generate co-existence. For instance, in order not to be either too many when the other species shrinks, or too few when the other species thrives, either predators, or preys, or both of them may decide to bind their reproduction rate to  the other species's success:

\begin{enumerate}
  \item Predators reproduce proportionally to the fraction of preys.  \label{cond:wolvesFraction}
  
  \item Preys  reproduce proportionally to the fraction of predators.  \label{cond:sheepFraction}
  
  \item Predators  reproduce proportionally to the fraction of preys and preys  reproduce proportionally to the fraction of predators.   \label{cond:sheepWolvesFraction}
\end{enumerate}

Behaviors~\ref{cond:wolvesFraction}, \ref{cond:sheepFraction} and~\ref{cond:sheepWolvesFraction} appear sensible but, surprisingly, none of them works very well. With~\ref{cond:wolvesFraction} the model ends up in $E_1$ $64.6\%$ of the times, in $E_2$ $12.3\%$ of the times and only $23.1\%$ of the times in $E_3$. This last figure sums a $13.6\%$ of cases where the population of predators oscillates around a fairly stable population of preys to a  $9.5\%$ where both populations oscillate. With~\ref{cond:sheepFraction} and~\ref{cond:sheepWolvesFraction} it is even worse, making the model reach $E_2$   $100\%$ and $99.6\%$ of the times, respectively. 

Insofar it concerns preys, this experiment  confirms Proposition~\ref{proposition:betterStupid}, namely that even a tiny bit of individual intelligence destroys collective intelligence. Predators' intelligence can have a positive impact on collective intelligence, but a $23.1\%$ of co-existence amounts to a modest success. Even more disturbing, this limited success disappears altogether if preys adopt the same  reasonable behavior.

\subsection{The Good KIDS}   \label{subsec:agree}

Let us  explore the possibility that, as expressed by Proposition~\ref{proposition:meta-cognition}, at least some individuals can envision the collective dynamics and attempt to  influence it by selecting some appropriate behavior. Specifically, let us suppose that either predators or preys are capable of meta-cognition, or both.

Let us suppose that the predators realize that, if there are too many of them, preys will disappear so in the end they  will also go extinct.  Therefore, they may collectively decide to stop reproducing  if their population becomes larger than the population of preys. In other words, predators  capable of meta-cognition steer the ecology towards co-existence by exerting double-loop learning (Proposition~\ref{proposition:meta-cognition_double-loop}).

Likewise, preys may collectively decide to stop reproducing  if their population becomes larger than the population of predators. Self-restraint could be sought, for instance, in order not to be too visible to predators. Preys capable of meta-cognition understand the system but cannot steer it towards the goal of co-existence, as it is typical of deutero-learning (Proposition~\ref{proposition:meta-cognition_deutero}); however, they can indirectly influence the system by exerting self-restraint.

Finally,  behaviors can be combined by assuming that both predators and preys stop reproducing if their own population becomes larger than the other one. This may result from some inter-species agreement as prescribed by Qualification~\ref{proposition:meta-cognition_agreement} or, equivalently, predators and preys may  make this move independently of one another.

\begin{enumerate}
\setcounter{enumi}{3}

  \item Predators stop reproducing if  their population becomes larger than the population of preys.  \label{cond:altruisticWolves}
  
  \item Preys stop reproducing if  their population becomes larger than the population of predators. \label{cond:altruisticSheep}
  
  \item Predators stop reproducing if  their population becomes larger than the population of preys and preys stop reproducing if  their population becomes larger than the population of predators. \label{cond:altruisticWolvesSheep}
\end{enumerate}

Note that since  \ref{cond:altruisticWolves}, \ref{cond:altruisticSheep} and~\ref{cond:altruisticWolvesSheep}  have been elaborated out of an understanding of global dynamics, they depend non-linearly on global thresholds. By contrast,  \ref{cond:wolvesFraction}, \ref{cond:sheepFraction} and~\ref{cond:sheepWolvesFraction} rely on shortsighted adaptation.

Rules  \ref{cond:altruisticWolves}, \ref{cond:altruisticSheep} and~\ref{cond:altruisticWolvesSheep}  are  remarkably similar to the strategies expressed by the mathematical continuous-time model~\ref{eq:lotka-volterra}. Indeed, assuming that  predators are willing to stop reproducing  if preys are too few (behavior~\ref{cond:altruisticWolves}) corresponds to a small coefficient~$d$  in eq.~\ref{eq:lotka-volterra}, whereas assuming that preys are prepared to stop reproducing  if predators are too few corresponds to a large coefficient~$b$.  Thus, at least two of the four coefficients of eqs.~\ref{eq:lotka-volterra} are being changed towards making the equilibrium  $E_3 = (c/d, a/b)$ globally  stable (see Appendix~\ref{app:lyapunov} for details).

Let us observe the results. If predators behave as in~\ref{cond:altruisticWolves}, $E_1$ is reached $27.2\%$ of the times, $E_2$ $3.1\%$ of the times, whereas~$E_3$ is reached~$69.7\%$ of the times. Thus, quite often this behavioral rule  makes predators and preys coexist. To a closer scrutiny, this successful~$69.7\%$ arises out of a~$0.8\%$ of outcomes where the two populations are roughly constant, a $0.3\%$ where the population of preys oscillates whereas predators do not, a~$38.9\%$ where the population of predators oscillates whereas preys do not and a $29.7\%$ where both populations oscillate.

However, if preys behave as in~\ref{cond:altruisticSheep} the outcome is radically different. In this case, neither $E_1$ (extinction of both predators and preys) nor $E_3$ (co-existence) ever occur, $E_2$ (exclusive survival of preys) happens $100.0\%$ of the times. Remarkably, preys' self-restraint turns to their own advantage. Note also that in this case, albeit only preys survive, they do not grow indefinitely as in the basic LV.

If both predators and preys exert meta-cognition~\ref{cond:altruisticWolvesSheep} the outcome is not a mixture of~\ref{cond:altruisticWolves} and~\ref{cond:altruisticSheep}, but rather equivalent to~\ref{cond:altruisticSheep}. Interestingly, in this context the indirect influence of deutero-learning (Proposition~\ref{proposition:meta-cognition_deutero}) proves stronger than the direct control enabled by double-loop learning  (Proposition~\ref{proposition:meta-cognition_double-loop}).

One may remark that case~\ref{cond:altruisticWolves} is remindful of the sort of relations that humans (the predators)  entertain with the  ecosystem (the preys). Humans are  predators who are capable of meta-cognition, and therefore they are in a position that could enable them to steer the ecosystem towards sustainable co-existence. Unfortunately, this requires predators to sacrifice their immediate self-interest in exchange for future advantages, a level of wisdom that may be difficult to attain.

Quite often, Nature endows preys with high reproduction rates in order to compensate for predation. However, cases~\ref{cond:altruisticSheep} and~\ref{cond:altruisticWolvesSheep} may exist in certain ecosystems where Nature selected preys for reproduction rates sufficiently low not to attract predators \cite{turner-75E}. Among humans,  one may notice that so-called ``primitive'' social groups eventually survived in environments so poor of natural resources to be ignored by more ``developed'' fellows. Or, public companies may hide profits in order to discourage hostile takeovers.

\subsection{The Perfect KIDS}   \label{subsec:perfectForesight}

By combining Propositions~\ref{proposition:prediction}  and~\ref{proposition:meta-cognition} one  obtains agents who have the ability to make  predictions based on a global understanding of the LV world that they inhabit. Specifically, let us ascribe to either predators, or preys, or both of them the capability of ``rational expectations'' enabled by: (\textit{i}) Knowledge of the model of the world as expressed by eqs.~\ref{eq:lotka-volterra}; (\textit{ii}) Capability to carry out any sort of computation, and (\textit{iii}) Availability of any required information.

Such agents may take the derivatives of eqs.~\ref{eq:lotka-volterra}, obtaining four equations that they can use to estimate coefficients $a$, $b$, $c$, $d$ and compute the LV-equilibrium $(c/d, \; a/b)$:

\begin{equation}        \label{eq:abcd}
  \left\{
  \begin{array}{rcl}
    a & = & (x \dot{x} \dot{y} + \dot{x}^{2} y - x \ddot{x} y) / x^2 \dot{y} \\
    b & = & (\dot{x}^2 - x \ddot{x}) / x^2 \dot{y} \\
    c & = & (x y \ddot{y} - x \dot{y}^2 - \dot{x} y \dot{y}) / \dot{x} y^2 \\
    d & = & (y \ddot{y} - \dot{y}^2) / \dot{x} y^2 \\
  \end{array}
  \right.
\end{equation}
where $a, b, c , d$ are the coefficients of eqs.~\ref{eq:lotka-volterra}.

Let us assume that predators and preys use these coefficients as follows (details in appendix~\ref{app:agent-based-LV}):

\begin{enumerate}
\setcounter{enumi}{6}

  \item Predators observe $x$, $y$, $\dot{x}$, $\dot{y}$, $\ddot{x}$, $\ddot{y}$, use eqs.~\ref{eq:abcd} to compute their equilibrium population $a/b$ and resize their population accordingly. \label{cond:wolvesPerfect}

  \item Preys observe $x$, $y$, $\dot{x}$, $\dot{y}$, $\ddot{x}$, $\ddot{y}$, use eqs.~\ref{eq:abcd} to compute their equilibrium population $c/d$ and resize their population accordingly. \label{cond:sheepPerfect}

  \item Both predators and preys observe $x$, $y$, $\dot{x}$, $\dot{y}$, $\ddot{x}$, $\ddot{y}$, use eqs.~\ref{eq:abcd} to compute  equilibrium  $E_3 =$ $(c/d, \; a/b)$ and resize their populations accordingly. \label{cond:wolvesSheepPerfect}
\end{enumerate}

A parameter had to be added in order to specify how much faster one population had to grow in order to safely assume that it would superseed the other one. I set this parameter to the default value of 100, supporting this choice with a sensitivity analysis reported in Appendix~\ref{app:sensitivity}.

The results are  discomforting. With~\ref{cond:wolvesPerfect}, $E_1$, $E_2$ and $E_3$ are reached $34.2\%$, $65.4\%$ and $0.4\%$ of the times. With~\ref{cond:sheepPerfect}, $E_1$, $E_2$ and $E_3$ are reached $39.6\%$, $60.2\%$ and $0.2\%$ of the times. With~\ref{cond:wolvesSheepPerfect}, $E_1$, $E_2$ and $E_3$ are reached $35.8\%$, $64.1\%$ and $0.1\%$ of the times. In all cases, co-existence is reached roughly just as often as in the basic case.

The reason is that the equilibrium   $E_3 =(c/d, \; a/b)$ is not structurally stable. An ABM introduces small deviations that easily take the system away from equilibrium, towards the  dynamics of the basic model. Indeed, Qualification~\ref{qualification:speed}) concerned stable equilibria generated by collective agreements.

\subsection{The Smart KISS of Simple KIDS}   \label{subsec:extrapolation}

Let us  consider the possibility that predators and preys make predictions by making use of linear extrapolation, which amounts to combining Proposition~\ref{proposition:prediction} with some intermediate level between KISS (Proposition~\ref{proposition:betterStupid}) and KIDS (Proposition~\ref{proposition:meta-cognition}). In particular, let us assume that either predators, or preys, or both of them reproduce out of extrapolation of the other species' reproductive trends.

Let us suppose that the ability to compute a derivative is employed for rather short-sighted ends. Specifically, let us assume that predators reproduce proportionately to the increase of  preys, and  let us assume that preys reproduce proportionately to the decrease of predators.

\begin{enumerate}
\setcounter{enumi}{9}

  \item Predators reproduce  with probability  proportional to the increase of the number of preys in the last simulation steps.  \label{cond:wolvesPredict}
  
  \item Preys reproduce with probability  proportional  to the decrease of the number of predators in the last simulation steps.  \label{cond:sheepPredict}
  
  \item Predators  reproduce with probability  proportional to the increase of the number of preys in the last simulation steps and preys reproduce with probability  proportional to the decrease of the number of predators in the last  simulation steps.  \label{cond:wolvesSheepPredict}
\end{enumerate}

Note that with rules~\ref{cond:wolvesPredict} and~\ref{cond:wolvesSheepPredict} predators do not reproduce if the population of preys did not increase. Similarly, with rules~\ref{cond:sheepPredict} and~\ref{cond:wolvesSheepPredict} preys do not reproduce if predators did not decrease. 

Similarly to\S~\ref{subsec:perfectForesight}, also in this case it is necessary to add a parameter in order to specify how many past steps the above variations are computed on. I set this parameter to the default value of 5 steps, supporting this choice with a sensitivity analysis reported in Appendix~\ref{app:sensitivity}.

Behavior~\ref{cond:wolvesPredict} is  effective in generating co-existence. With~\ref{cond:wolvesPredict},  generalized extinction~$E_1$ never occurs. Preys-dominated equilibrium $E_2$  occurs on average $0.1\%$ of the times. Most importantly, $E_3$ was reached $99.9\%$ of the times. Out of this $99.9\%$,   $7.3\%$ of the times neither population oscillated, $1.8\%$ of the times the population of preys oscillated around a roughly constant population of predators whereas  $77.2\%$ of the times  the opposite occurred and, finally, in the remaining $13.6\%$ both populations oscillated. In a nutshell, preys and predators coexist and, most often, the population of predators oscillates while the population of preys remains roughly constant.

In the business cycle interpretation of the LV model, predators are workers who ask for higher salaries when unemployment is low. Amid many  specifications of this relation, direct proportionality between the variation of unemployment and the variation of wages --- essentially, behavioral rule~\ref{cond:wolvesPredict} --- provides the best fit to empirical data \cite{gordon-89AER} \cite{skovranek-19M}. Note also that trade unions that understand this dynamics and are willing to moderate their requests in order to smooth the business cycle are often put aside by more aggressive competitors, which is an instance of deutero-learning (Proposition~\ref{proposition:meta-cognition_deutero}).

Also in the opposite case~\ref{cond:sheepPredict} co-existence of preys and predators $E_3$ is easily reached, namely $100.0\%$ of times. However, in this case with a frequency of $48.4\%$ the two populations remain roughly constant,  $24.6\%$ of times the population of preys oscillated around a nearly constant population of predators,  $13.7\%$ of the times the population of predators oscillated around a nearly constant population of preys, whereas both populations oscillated in the remaining $13.3\%$ of the times. While in case~\ref{cond:wolvesPredict} the population of predators most often oscillates, with~\ref{cond:sheepPredict} predators rarely oscillate whereas preys either oscillate or stay roughly constant.

In the business cycle interpretation of the LV model, preys are businesses that invest (the derivative of capital) proportionately to the variation of income  \cite{samuelson-39RES} \cite{gabisch-lorenz-89}, which corresponds to behavioral rule~\ref{cond:sheepPredict}. According to the above simulations either predators (employees) or preys (businesses) can attain co-existence by making predictions, which is an instance of double-loop learning (Proposition~\ref{proposition:meta-cognition_double-loop}), though in practice these behaviors are combined into~\ref{cond:wolvesSheepPredict}.

Interestingly, the mixed arrangement~\ref{cond:wolvesSheepPredict} is not some sort of average between~\ref{cond:wolvesPredict} and~\ref{cond:sheepPredict} but rather exhibits a dynamics on its own. With~\ref{cond:wolvesSheepPredict}, $E_1$ is reached $40.4\%$, $E_2$  $0.1\%$, and $E_3$ is reached $59.5\%$ of the times. Out of this  $59.5\%$, $55.9\%$ of the times both populations oscillate, $3.3\%$  only the population of preys oscillates,  $0.2\%$ only predators oscillate and, finally, in the remaining $0.1\%$ neither population oscillates. Thus, co-existence is reached quite often, and  most of the times it is reached with both populations oscillating.

\begin{figure}
\center
\fbox{\resizebox*{0.8\textwidth}{!}{\includegraphics{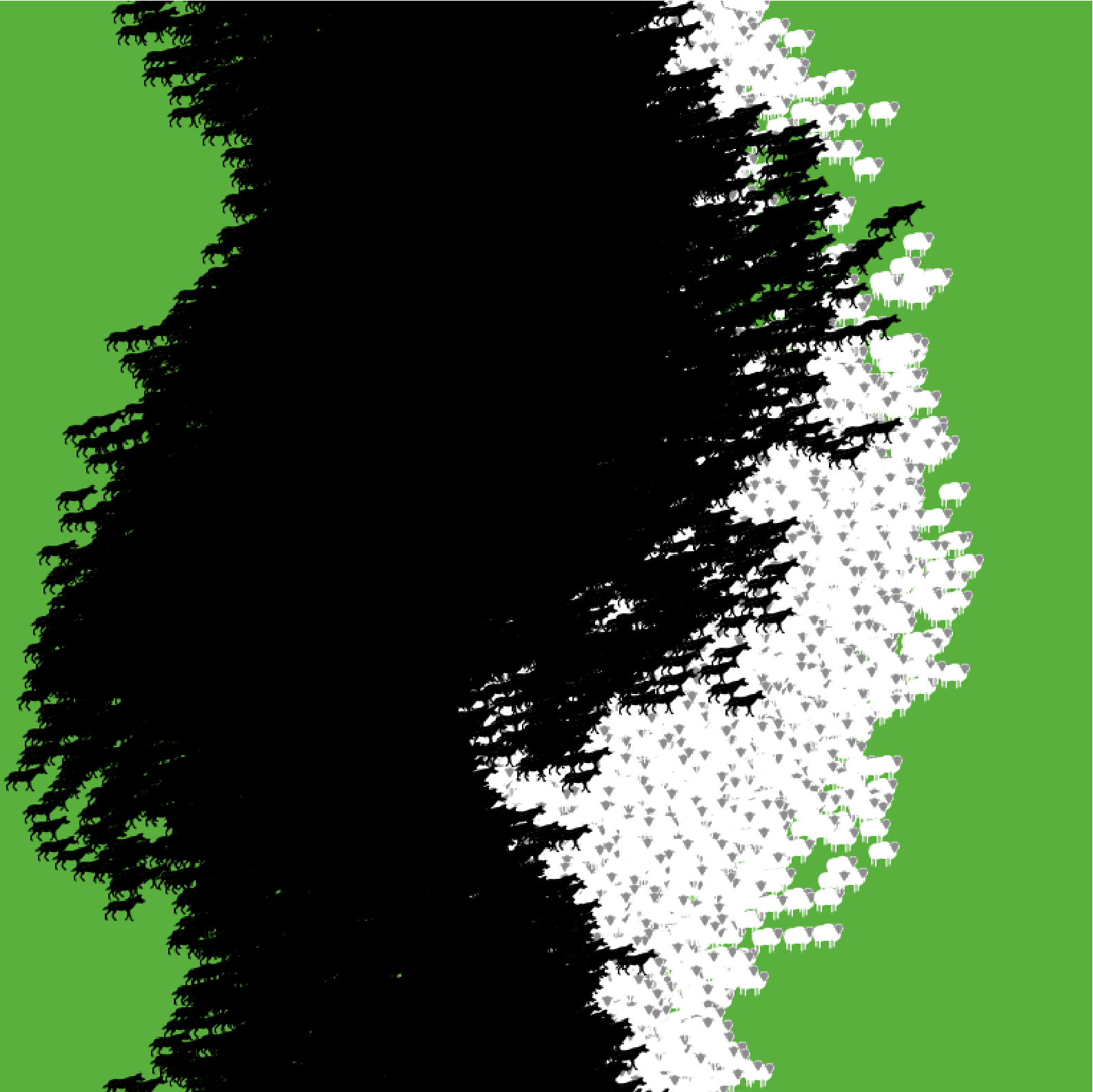}}}
\caption{The most common outcome of case~\ref{cond:wolvesSheepPredict}. Both populations grow while  predators (black) are chasing preys (white). Co-existence is achieved, but while both populations are growing.} \label{fig:chase}
\end{figure}

However, this sort of co-existence is quite remarkable because, differently from all previous cases where $E_3$ occurred, both populations grow. A large herd of preys moves in  artificial space, hunted by an even larger herd of predators. Preys reproduce while running away from predators, predators reproduce while chasing preys generating the two moving columns shown  in Figure~\ref{fig:chase}.

\begin{figure}
\center
\fbox{\resizebox*{0.8\textwidth}{!}{\includegraphics{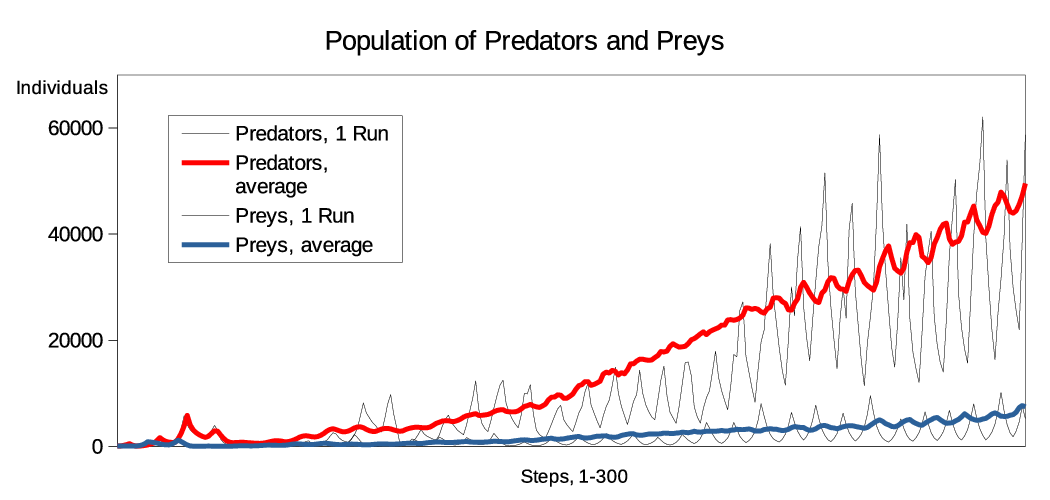}}}
\caption{The oscillations of the population of predators, averaged over the 595 (out of 1,000) time series where $E_3$ obtains (thick blue line) plotted against one single time series selected for being closest to the mean (thin black line). The oscillations of the population of preys, averaged over the 595 time series where $E_3$ obtains (thick green line) plotted against one single time series selected for being closest to the mean (thin dotted line). On average, the population of predators grows from 50 to 47,817.03  individuals whereas the population of preys grows from 100 to 6,544.96 individuals along 500 time steps.} \label{fig:timeSeries}
\end{figure}

Figure~\ref{fig:timeSeries} illustrates the average of the 595 time series that reach co-existence ($E_3$), plotted against one of them in order to show individual variability. Preys grow 60-fold, predators grow almost 1,000-fold over 500 time steps. This sort of explosive Lotka Volterra (ELV) differs from LV versions where predators and preys grow out of symbiotic relations \cite{fort-24I} because it originates from competitive, purely self-interested behavior.

In the business cycle interpretation it would be unrealistic to assume that only predators or only preys conform to the model. Case~\ref{cond:wolvesSheepPredict} is relevant, rather than either~\ref{cond:wolvesPredict} or~\ref{cond:sheepPredict}. This implies that ELV obtains, rather than LV, and indeed, the business cycle is inseparable from economic growth.

\section{Conclusions}  \label{sec:conclusions}

The exploration of the capability of individual algorithms more sophisticated than those expressed by the basic LV to generate collectively more ``intelligent'' behavior was in general unsuccessful except for two classes of algorithms, namely those that express meta-cognition (see \S~\ref{subsec:agree}) and those  that express their capability to make predictions (see \S~\ref{subsec:extrapolation}), respectively. In a landscape that generally favors KISS over KIDS, these capabilities stand out as  game-changers.

The importance of meta-cognition may appear obvious, but its consequences --- at least in the LV --- are definitely not. In particular, it is not obvious that preys that must content themselves with deutero-learning  may end up with pushing predators to extinction. However, if ``extinction'' is replaced with the less ambitious objective of diverting the attention of predators towards other preys, then the advantage of reducing visibility becomes obvious.

Other prototypes of collective behavior that are likely be strongly affected by meta-cognition may include the Spacial Segregation Model, as well as the aforementioned Traveler's Dilemma. However, these models do not exhibit a sharp separation between agents that are capable of double-loop learning from agents that limit themselves to deutero-learning. In these models, all agents exert double-loop learning so the outcome is likely to be less surprising.

The ability to make predictions appears to have an even greater impact on global dynamics than meta-cognition. In particular, the ELV appears to capture the explosive dynamics of capitalistic economies otherwise ignored by equilibrium models. It deserves an analytical description grounded on economic theories and empirical stylized facts, that unfortunately exceeds the scope of this work.

Similarly, to meta-cognition, one may ask whether prediction generates similarly remarkable aggregate dynamics in other models. Furthermore, one may ask whether higher-order derivatives at the individual level are similarly important in generating interesting collective dynamics. 

Tentatively, I am inclined to answer the first question with a ``Yes,''  the second one with a ``No.'' One the one hand, prediction has a substantial impact on many models of collective behavior, ranging from the Minority Game to The Prisoner's Dilemma. On the other hand, if interacting individuals compute first-order derivatives, then the system as a whole is capable of computing derivatives of any order.

\section*{Conflicts of interest and other legal disclaimers}

The author declares not to have any conflict of interest whatsoever with the organizations, institutions and persons directly or indirectly mentioned in this paper. The author did not receive any funding to carry out this research.

All figures and tables have been generated by the authors. No empirical data have been used. No experiment involving humans or other animals was carried out.

\appendix

\section{Lyapunov Theorem  on the Lotka-Volterra Model}   \label{app:lyapunov}

The LV eqs.~(\ref{eq:lotka-volterra}) generate stable cycles. Global asymptotic stability of a cycle can be investigated by means of the Lyapunov Theorem.

Consider a nonlinear two-dimensional dynamical system:

\begin{displaymath}  
  \left\{
  \begin{array}{rcl}
    \dot{x} & = & f(x, y) \\
    \dot{y} & = & g(x, y)
  \end{array}
  \right.
\end{displaymath}
where $x, y \in \Re$ and $f, g \in \Re^2 \mapsto \Re$.

Without any loss of generality, let us assume that this system has an equilibrium point at $E = (x^*, y^*)$. The Lyapunov Theorem states that if a function $V: \Re^2 \mapsto \Re$ exists, such that:

\begin{enumerate}
  \item $V(x, y) = 0$ if and only if $(x, y) = (x^*, y^*)$;  \label{cond:zero}
  
  \item $V(x, y) > 0$ if and only if $(x, y) \neq (x^*, y^*)$;  \label{cond:GTzero}
  
  \item $\dot{V}(x, y) = \frac{d}{dt} V(x,y) = \frac{\partial V}{\partial x} f(x, y) +  \frac{\partial V}{\partial y} g(x, y)  \leq 0$ for $(x, y) \neq (x^*, y^*)$;   \label{cond:derivative}
\end{enumerate}
then   $(x^*, y^*)$ is a stable equilibrium point. Function $V(x, y)$ is  a \emph{Lyapunov function}.

The Lotka-Volterra model~(\ref{eq:lotka-volterra}) has an equilibrium  at $(x^*, y^*) = (c/d, \, a/b)$. Let us investigate its global stability by means of the following Lyapunov function  \cite{takeuchi-96}:

\begin{displaymath}
  V(x, y) = d (x - x^* \ln x) + b (y - y^* \ln y) + [d (x^* - x^* \ln x^*) + b (y^* - y^* \ln y^*)]
\end{displaymath}
which we may also write as $V(x(t), y(t)) = d (x - c/d \ln x) + b (y - a/b \ln y) + const$.

It is obviously $V(x^*, y^*) = 0$. Thus, condition~(\ref{cond:zero}) is satisfied.

Let us check whether $V(x, y) >0$. Let us consider the first term, namely, $d (x - x^* \ln x)$. It is $d (x - x^* \ln x) > 0$ if $d x - c \ln x > 0$ $\rightarrow$ $d x > c \ln x$ $\rightarrow$ $x > c/d \ln x$ $\rightarrow$ $e^x > e^{c/d \ln x}$ $\rightarrow$ $e^x > x^{c/d}$ which is always true because the exponential function yields greater values than the power function. The third term, $d (x^* - x^* \ln x^*)$, makes sure that the combination of the first and the third term starts to yield positive values just as soon as $V$ leaves equilibrium $(x^*, y^*)$. Likewise, $e^y > y^{a/b}$ is always true because the exponential function yields greater values than the power function and the fourth term $b (y^* - y^* \ln y^*)$ ensures that this only happens outside $(x^*, y^*)$. Thus, condition~(\ref{cond:GTzero}) is satisfied.

Let us turn to the third condition. It is $\dot{V}$ $ =$ $(d - d \frac{x^*}{x}) (a x - bxy) + (b - b \frac{y^*}{y}) (-c y + d xy)$ $=$ 
$ad (x - x^*) - bc (y- y^*) + bd (x^*y - y^* x)$. Obviously, $\dot{V}(x^*, y^*) = 0$. Let us investigate which parameters make for $\dot{V}(x, y) < 0$ at $(x, y) \neq (x^*, y^*)$.

Let us explore the sign of $\partial \dot{V} / \partial x = ad -bd y^*$ and $\partial \dot{V} / \partial y = -bc +bd x^*$. $\dot V$ is negative if:

\begin{displaymath}
\left \{
  \begin{array}{rcl}
    ad - bd y^* & < & 0          \\
    -bc + bd x^* & < & 0  
  \end{array}
  \right.
\end{displaymath}

We obtain that $\dot V$ is negative if:

\begin{displaymath}
\left \{
  \begin{array}{rcl}
    a & < & b y^*          \\
    c & > & d x^*  
  \end{array}
  \right.
\end{displaymath}

This means that:

\begin{itemize}
  \item The smaller $a$, the more likely that $\dot{V} <0$;
  
  \item The greater $b$, the more likely that $\dot{V} <0$.
  
  \item The greater $c$, the more likely that $\dot{V} <0$.
  
  \item The smaller $d$, the more likely that $\dot{V} <0$.
\end{itemize}

Thus,  with sufficiently small~$a$, large~$b$, large~$c$ and small~$d$ the equilibrium  $(c/d, a/b)$ is globally  stable.

\section{An Agent-Based Lotka-Volterra}~\footnote{This model was jointly developed by Andrea Policarpi and Guido Fioretti. Our model is freely available on \emph{CoMSES} under GPL 3.0.}   \label{app:agent-based-LV}

Since we built our model out of Wilensky and Reisman's model \cite{wilensky-97} \cite{wilensky-reisman-06CI}, our code keeps  naming  predators as ``Wolves'' and preys as ``Sheep,'' respectively. Specifically, we derived our model from the \textbf{sheep-wolves} version which lacks the agent ``Grass'' where we eliminated energy halving whenever preys reproduce. Correspondingly, we also eliminated the parameters \textbf{grass-regrowth-time},  \textbf{sheep-gain-from-food}. By contrast, parameters  \textbf{initial-sheep},  \textbf{initial-wolves},  \textbf{sheep-reproduce},  \textbf{wolf-reproduce} and \textbf{wolves-gain-from-food} carry on to our model.

We kept the five parameters that we mutuated from Wilensky and Reisman's model at their original default values, whereas we used  switches to select among behavioral configurations in~\S~\ref{sec:prey-predator}. We introduced two new parameters in order to explore the behavior configurations of \S~\ref{subsec:perfectForesight} and \S~\ref{subsec:extrapolation}, which we named  \textbf{how-much-faster} and   \textbf{Timespan}, respectively.  We kept these two additional parameters  at reasonable base values throughout our simulations, carrying out a sensitivity analysis whose results are reported in Appendix~\ref{app:sensitivity}. The ensuing Table~\ref{tab:parameters} summarizes the parameter values of our model for each behavior configuration, as well as the switches that we used to select among them.

\begin{sidewaystable}
\begin{tabular}{|l|c|c|c|c|c|c|c|c|c|c|c|c|c|}
\cline{2-14}
 \multicolumn{1}{c|}{} & \multicolumn{13}{|c|}{Behavior Configuration} \\
\hline
 Parameter Name & 0 & \ref{cond:wolvesFraction} & \ref{cond:sheepFraction} & \ref{cond:sheepWolvesFraction} & \ref{cond:altruisticWolves} & \ref{cond:altruisticSheep} & \ref{cond:altruisticWolvesSheep} & \ref{cond:wolvesPerfect} & \ref{cond:sheepPerfect} & \ref{cond:wolvesSheepPerfect} & \ref{cond:wolvesPredict} & \ref{cond:sheepPredict} & \ref{cond:wolvesSheepPredict} \\
\hline
 \textbf{initial-sheep} & 100 & 100 & 100 & 100 & 100 & 100 & 100 & 100 & 100 & 100 & 100 & 100 & 100 \\
 \textbf{initial-wolves} & 50 & 50 & 50 & 50 & 50 & 50 & 50 & 50 & 50 & 50 & 50 & 50 & 50 \\
 \textbf{sheep-reproduce} & 4 & 4 & 4 & 4 & 4 & 4 & 4 & 4 & 4 & 4 & 4 & 4 & 4 \\
 \textbf{wolf-reproduce} & 5 & 5 & 5 & 5 & 5 & 5 & 5 & 5 & 5 & 5 & 5 & 5 & 5 \\
 \textbf{wolves-gain-from-food} & 20 & 20 & 20 & 20 & 20 & 20 & 20 & 20 & 20 & 20 & 20 & 20 & 20 \\
\hline
 \textbf{Wolves-repr-by-frac-sheep} & Off & \cellcolor{yellow} On & Off & \cellcolor{yellow} On & Off & Off & Off & Off & Off & Off & Off & Off & Off \\
 \textbf{Sheep-repr-by-frac-wolves} & Off & Off & \cellcolor{yellow} On & \cellcolor{yellow} On & Off & Off & Off & Off & Off & Off & Off & Off & Off \\
\hline
 \textbf{Stop-rep-if-wolves$>$sheep} & Off & Off & Off & Off & \cellcolor{yellow} On & Off & \cellcolor{yellow} On & Off & Off & Off & Off & Off & Off \\
 \textbf{Stop-rep-if-sheep$>$wolves} & Off & Off & Off & Off & Off & \cellcolor{yellow} On & \cellcolor{yellow} On & Off & Off & Off & Off & Off & Off \\
\hline
 \textbf{Wolves-perfect-foresight} & Off & Off & Off & Off & Off & Off & Off & \cellcolor{yellow} On & Off & \cellcolor{yellow} On & Off & Off & Off \\
 \textbf{Sheep-perfect-foresight} & Off & Off & Off & Off & Off & Off & Off & Off & \cellcolor{yellow} On & \cellcolor{yellow} On & Off & Off & Off \\
 \textbf{how-much-faster} & 10 & 10 & 10 & 10 & 10 & 10 & 10 & 10 & 10 & 10 & 10 & 10 & 10 \\
\hline
 \textbf{Wolves-extrapolate} & Off & Off & Off & Off & Off & Off & Off & Off & Off & Off & \cellcolor{yellow} On & Off & \cellcolor{yellow} On \\
 \textbf{Sheep-extrapolate} & Off & Off & Off & Off & Off & Off & Off & Off & Off & Off & Off & \cellcolor{yellow} On & \cellcolor{yellow} On \\
 \textbf{Timespan} & 5 & 5 & 5 & 5 & 5 & 5 & 5 & 5 & 5 & 5 & 5 & 5 & 5 \\
\hline
 \textbf{Simulation Length}  & 500 & 500 & 500 & 500 & 500 & 500 & 500 & 500 & 500 & 500 & 500 & 500 & 500 \\
 \textbf{max-wolves}  & 500k & 500k & 500k & 500k & 500k & 500k & 500k & 500k & 500k & 500k & 500k & 500k & $\:\:$1M$\:\:$ \\
 \textbf{max-sheep}  & 500k & 500k & 500k & 500k & 500k & 500k & 500k & 500k & 500k & 500K & 500k & 500k & $\:\:$1M$\:\:$ \\
 \textbf{Transitory}  & 50 & 50 & 50 & 50 & 50 & 50 & 50 & 50 & 50 & 50 & 50 & 50 & 50 \\
 \textbf{Benchmarking}  & 100 & 100 & 100 & 100 & 100 & 100 & 100 & 100 & 100 & 100 & 100 & 100 & 100 \\
\hline
\end{tabular}
\caption{The switches and parameters of our model, for each behavior configuration. Switches that are ``On'' for a specific configuration are highlighted in yellow. The basic Wilensky model is denoted as the zero-configuration, the remaining ones are numbered as in~\S~\ref{sec:prey-predator}. In the top five lines, the five parameters inherited from Wilensky's model. Subsequently, the two parameters relative to configurations \ref{cond:wolvesFraction}, \ref{cond:sheepFraction}, \ref{cond:sheepWolvesFraction} and the two parameters relative to configurations \ref{cond:altruisticWolves}, \ref{cond:altruisticSheep}, \ref{cond:altruisticWolvesSheep}, respectively. Then the three parameters that are relative to configurations \ref{cond:wolvesPerfect}, \ref{cond:sheepPerfect}, \ref{cond:wolvesSheepPerfect} and \ref{cond:wolvesPredict}, \ref{cond:sheepPredict}, \ref{cond:wolvesSheepPredict}, respectively. In the bottom lines,  five parameters that are employed to run and observe the model.}  \label{tab:parameters}
\end{sidewaystable}

By tuning parameters \textbf{Simulation Length}, \textbf{max-wolves} and \textbf{max-sheep} our model runs until either the maximum number of steps is reached, or both wolves and sheep have died out, or either population has reached the maximum allowed. Specifically, we ran the model for 500 steps and allowed a maximum of 500,000 wolves and 500,000 sheep, respectively. These values were chosen because 500 steps were more than enough to observe results, and because the 500,000 threshold was reached only when either wolves or sheep were the only surviving species.

In order to observe outputs we also introduced a parameter \textbf{Transitory} and a parameter \textbf{Benchmarking} that we set at 50 and 100 steps, respectively. Between  \textbf{Transitory} and  \textbf{Benchmarking}  the model determines which dynamics count as oscillations. At each  step within this interval, the minimum and maximum number of sheep and wolves that have been ever attained are updated. Let us call them $MINs$, $MAXS$, $MINW$ AND $MAXW$, respectively. Sheep oscillations are counted if $(MAXS - MINS) / 2 \geq 0.2   (MAXS + MINS) / 2$. If this condition is satisfied, one oscillation occurs if the number of sheep is greater or equal to $( MAXS + MINS ) / 2 + 0.5 ( MAXS - MINS ) / 2 )$ and, subsequently, the number of wolves is smaller or equal to $( MAXW + MINW ) / 2 ) - 0.5 ( MAXW - MINW ) / 2$. Wolves oscillations are counted in a similar way. Also the possibility of perfect foresight is only considered after  \textbf{Benchmarking} steps have elapsed.

The parameters that have been introduced in order to run the model and obtain outputs are listed in the lower portion of Table~\ref{tab:parameters}.  Since we are dealing with a simulation model that must be necessarily stopped after a finite number of steps, we eventually obtain outcomes that exhibit richer details than the mathematically defined  $E_1$, $E_2$ and~$E_3$. We mapped simulation outcomes onto $E_1$, $E_2$ and~$E_3$ as follows:

\begin{enumerate}
\setcounter{enumi}{-1}

\item When the simulation is stopped, neither sheep nor wolves exist. This case corresponds to $E_1$.

\item When the simulation is stopped, only wolves exist. Since this is untenable in the long run, we assumed that wolves would go extinct as well. Thus, also this case corresponds to $E_1$.

\item When the simulation is stopped, only sheep exist. This case corresponds to $E_2$.

\item Both sheep and wolves exist when the simulation is stopped, and no oscillation has been observed. We classified this case as $E_3$.

\item Both sheep and wolves exist when the simulation is stopped. The number of sheep has been oscillating, but no oscillation of the number of wolves has been observed. We classified this case as $E_3$.

\item Both sheep and wolves exist when the simulation is stopped. The number of wolves has been oscillating, but no oscillation of the number of sheep has been observed. We classified this case as $E_3$.

\item Both sheep and wolves exist when the simulation is stopped. Both the number of sheep and the number of wolves have been oscillating. We classified this case as $E_3$.

\item The simulation is stopped because the number of sheep reached a threshold imposed by available computational power. When the simulation was stopped the population of sheep was growing faster than the population of wolves. By extrapolating a future where sheep continue to grow whereas wolves become negligible and finally die out we classified this case as $E_2$.

\item The simulation is stopped because the number of sheep reached a threshold imposed by available computational power. Contrary to the previous case, when the simulation was stopped the population of wolves was growing faster than the population of sheep. By extrapolating a future where wolves dominate the scene, eat all the sheep and finally go extinct we classified this case as $E_1$.

\item The simulation is stopped because the number of wolves reached a threshold imposed by available computational power. When the simulation was stopped the population of wolves was growing faster than the population of sheep. By extrapolating a future where wolves dominate the scene, eat all the sheep and finally go extinct we classified this case as $E_1$.

\item The simulation is stopped because the number of wolves reached a threshold imposed by available computational power. Contrary to the previous case, when the simulation was stopped the population of sheep was growing faster than the population of wolves. By extrapolating a future where sheep continue to grow whereas wolves become negligible and finally die out we classified this case as $E_2$.

\end{enumerate}

Our \href{www.comses.net/codebases/0eada5b3-3d18-4fc6-92af-841ff0971d28/}{extended version} of the Lotka-Volterra model is available on \href{https://www.comses.net/}{CoMSES}. Further documentation is provided alongside with the code.

\section{Sensitivity Analysis}   \label{app:sensitivity}

I carried out a sensitivity analysis on the two parameters that I added to Wilensky's model \cite{wilensky-97} \cite{wilensky-reisman-06CI}, namely \textbf{how-much-faster} and \textbf{Timespan}. Since  \textbf{how-much-faster} regulates the difference of wolves and sheep growth rates  that activates their perfect foresight, I analyzed its impact when  behavioral rules~(7), (8) and~(9) are adopted. Likewise, since \textbf{Timespan} regulates the time interval over which extrapolations are made, I analyzed its impact when behavioral rules~\ref{cond:wolvesPredict}, \ref{cond:sheepPredict} and~\ref{cond:wolvesSheepPredict} are adopted.

In both cases I explored the consequences  on  $E_1$, $E_2$ and $E_3$ of  $40\%$ increments and decrements of these parameters with respect to their base values. For  \textbf{how-much-faster} this meant exploring the consequences of decreasing it to 60 and increasing it to 140 from its base value of 100.  For  \textbf{Timespan} this meant exploring the consequences of decreasing it to 3 and increasing it to 7 from its base value of 5. In both cases I measured the occurrence of $E_1$, $E_2$ and $E_3$ over 1,000 runs.

Table~\ref{tab:how-much-faster} reports the outcomes of the sensitivity analysis for parameter \textbf{how-much-faster}. The outcomes with this parameter at its base value are reported in the three central columns, whereas the outcomes when this parameter is decreased or increased by $40\%$ are reported in the three columns on their left and their right, respectively. It appears that variations of this parameter have no impact whatsoever.

\begin{table}
\begin{tabular}{c|r|r|r|r|r|r|r|r|r|}
\cline{2-10}
 &  \multicolumn{3}{c}{} & \multicolumn{3}{c}{\textbf{how-much-faster}}  & \multicolumn{3}{c|}{}  \\
 \cline{2-10}
 & \multicolumn{3}{c|}{60} & \multicolumn{3}{c|}{100} & \multicolumn{3}{c|}{140}  \\
\cline{2-10}
 &  \multicolumn{3}{c}{} & \multicolumn{3}{c}{\textbf{Behavioral Rules}}  & \multicolumn{3}{c|}{}  \\
 \cline{2-10}
 & \multicolumn{1}{c|}{\ref{cond:wolvesPerfect}} & \multicolumn{1}{c|}{\ref{cond:sheepPerfect}} & \multicolumn{1}{c|}{\ref{cond:wolvesSheepPerfect}} & \multicolumn{1}{c|}{\ref{cond:wolvesPerfect}} & \multicolumn{1}{c|}{\ref{cond:sheepPerfect}} & \multicolumn{1}{c|}{\ref{cond:wolvesSheepPerfect}} & \multicolumn{1}{c|}{\ref{cond:wolvesPerfect}} & \multicolumn{1}{c|}{\ref{cond:sheepPerfect}} & \multicolumn{1}{c|}{\ref{cond:wolvesSheepPerfect}}   \\
\hline
\multicolumn{1}{|c|}{$E_1$} & 0 & 1000 & 1000 & 0 & 1000 & 1000 & 0 & 1000 & 1000 \\
\multicolumn{1}{|c|}{$E_2$} & 1000 & 0 & 0 & 1000 & 0 & 0 & 1000 & 0 & 0 \\
\multicolumn{1}{|c|}{$E_3$} & 0 & 0 & 0 & 0 & 0 & 0 & 0 & 0 & 0  \\
\hline
\end{tabular}
\caption{The number of aggregate outcomes $E_1$, $E_2$ and $E_3$ with behavioral rules~\ref{cond:wolvesPerfect}, \ref{cond:sheepPerfect} and~\ref{cond:wolvesSheepPerfect} when \textbf{how-much-faster}  is decreased by $40\%$ from  100 to 60 and increased by $40\%$ from 100 to 140, respectively. All other parameters at their base values.}  \label{tab:how-much-faster}
\end{table}

Likewise, Table~\ref{tab:Timespan} reports the outcomes of the sensitivity analysis for parameter \textbf{Timespan}. The three central columns report outcomes with this parameter at its base value, whereas the three columns on their left and their right report the outcomes when this parameter is decreased by $40\%$ and increased by $40\%$, respectively.

It appears that  behavioral rules~\ref{cond:wolvesPredict} and~\ref{cond:sheepPredict} --- either predators or preys predict, not both of them --- are largely   sensitive to variations of $\mathbf{Timespan}$.

By contrast,  behavioral rule~\ref{cond:wolvesSheepPredict} --- both predators and preys predict ---  is sensitive to both lower and greater values of  $\mathbf{Timespan}$. Specifically,  it appears that  $\mathbf{Timespan}=5$ corresponds to a  minimum of $E_1$ and a maximum of $E_2$ and $E_3$.

\begin{table}
\begin{tabular}{c|r|r|r|r|r|r|r|r|r|}
\cline{2-10}
 &  \multicolumn{3}{c}{} & \multicolumn{3}{c}{\textbf{Timespan}}  & \multicolumn{3}{c|}{} \\
 \cline{2-10}
 & \multicolumn{3}{c|}{3} & \multicolumn{3}{c|}{5} & \multicolumn{3}{c|}{7}  \\
\cline{2-10}
 &  \multicolumn{3}{c}{} & \multicolumn{3}{c}{\textbf{Behavioral Rules}}  & \multicolumn{3}{c|}{}  \\
 \cline{2-10}
 & \multicolumn{1}{c|}{\ref{cond:wolvesPredict}} & \multicolumn{1}{c|}{\ref{cond:sheepPredict}} & \multicolumn{1}{c|}{\ref{cond:wolvesSheepPredict}} & \multicolumn{1}{c|}{\ref{cond:wolvesPredict}} & \multicolumn{1}{c|}{\ref{cond:sheepPredict}} & \multicolumn{1}{c|}{\ref{cond:wolvesSheepPredict}} & \multicolumn{1}{c|}{\ref{cond:wolvesPredict}} & \multicolumn{1}{c|}{\ref{cond:sheepPredict}} & \multicolumn{1}{c|}{\ref{cond:wolvesSheepPredict}}   \\
\hline
\multicolumn{1}{|c|}{$E_1$} & 0 & 0 & 823 & 1 & 0 & 143 & 0 & 0 & 458 \\
\multicolumn{1}{|c|}{$E_2$} & 0 & 0 & 11 & 0 & 0 & 230 & 0 & 0 & 53 \\
\multicolumn{1}{|c|}{$E_3$} & 1000 & 1000 & $\;\;$166 & $\;\;$999 & 1000 & $\;\;$627 & 1000 & 1000 & $\;\;$489  \\
\hline
\end{tabular} 
\caption{The number of aggregate outcomes $E_1$, $E_2$ and $E_3$ with behavioral rules~(10), \ref{cond:sheepPredict} and~(12) when \textbf{Timespan}  is decreased by $40\%$ from  5 to 3 and increased by $40\%$ from 5 to 7, respectively. All other parameters at their base values.}  \label{tab:Timespan}
\end{table}

\bibliographystyle{plain}
\bibliography{../../references}

\begin{thebibliography}{100}

\bibitem{argyris-03OS}
Chris Argyris.
\newblock A life full of learning.
\newblock {\em Organization Studies}, 24(7):1178--1192, 2003.

\bibitem{argyris-schon-78}
Chris Argyris and Donald~A. Sch\"on.
\newblock {\em Organizational Learning: A theory of action perspective}.
\newblock Addison-Wesley Publishing Company, Redwood City, 1978.

\bibitem{argyris-schon-96}
Chris Argyris and Donald~A. Sch\"on.
\newblock {\em Organizational Learning II: Theory, method, and practice}.
\newblock Addison-Wesley Publishing Company, Reading (MS), 1996.

\bibitem{arthur-aymansmith-01AMJ}
Jeffrey~B. Arthur and Lynda Aiman-Smith.
\newblock Gainsharing and organizational learning: An analysis of employee
  suggestions over time.
\newblock {\em Academy of Management Journal}, 44(4):737--754, 2001.

\bibitem{axelrod-84}
Robert Axelrod.
\newblock {\em The Evolution of Cooperation}.
\newblock Basic Books, New York, 1984.

\bibitem{bahar-mao-04JMAA}
Arifah Bahar and Xuerong Mao.
\newblock Stochastic delay lotka-volterra model.
\newblock {\em Journal of Mathematical Analysis and Applications},
  292(2):364--380, 2004.

\bibitem{basu-94AER}
Kaushik Basu.
\newblock The traveler's dilemma: Paradoxes of rationality in game theory.
\newblock {\em American Economic Review}, 84(2):391--395, 1994.

\bibitem{basu-07SA}
Kaushik Basu.
\newblock The traveler's dilemma.
\newblock {\em Scientific American}, 296(June):90--95, 2007.

\bibitem{bateson-58}
Gregory Bateson.
\newblock {\em Naven}.
\newblock Stanford University Press, Stanford, 1958.

\bibitem{bateson-72}
Gregory Bateson.
\newblock {\em Steps to an Ecology of Mind}.
\newblock Ballantine Books, New York, 1972.

\bibitem{beenhakker-huguenard-09N}
Mark~P. Beenhakker and John~R. Huguenard.
\newblock Neurons that fire together also conspire together: Is normal sleep
  circuitry hijacked to generate epilepsy?
\newblock {\em Neuron}, 62(June):612--632, 2009.

\bibitem{bhargava-89TFSC}
Subhash~C. Bhargava.
\newblock Generalized lotka-volterra equations and the mechanism of
  technological substitution.
\newblock {\em Technological Forecasting and Social Change}, 35(4):319--326,
  1989.

\bibitem{bion-61}
Wilfred~R. Bion.
\newblock {\em Experiences in Groups}.
\newblock Tavistock Publications, London, 1961.

\bibitem{carmichael-hadzikadic-13ACS}
Ted Carmichael and Mirsad Hadzikadic.
\newblock Emergent features in a general food web simulation: Lotka-volterra,
  gause's law, and the paradox of enrichment.
\newblock {\em Advances in Complex Systems}, 16(8):1350014, 2013.

\bibitem{choi-97NDPLS}
Chang-Hyeong Choi.
\newblock Generalizations of the lotka-volterra population ecology model:
  Theory, simulation, and applications.
\newblock {\em Nonlinear Dynamics, Psychology and Life Sciences},
  4(2):263--273, 1997.

\bibitem{christodoulakis-16DPE}
Nicos Christodoulakis.
\newblock Conflict dynamics and costs in the greek civil war 1946–1949.
\newblock {\em Defence and Peace Economics}, 27(5):688--717, 2016.

\bibitem{clark-93}
Andy Clark.
\newblock {\em Associative Engines: Connectionism, Concepts, and
  Representational Change}.
\newblock The MIT Press, Cambridge, 1993.

\bibitem{clark-chalmers-98A}
Andy Clark and David~J. Chalmers.
\newblock The extended mind.
\newblock {\em Analysis}, 58(1):7--19, 1998.

\bibitem{nunesdecastro-vonzuben-dedeus-03N}
Leandro~Nunes de~Castro, Fernando J.~Von Zuben, and Jr. Get\'ulio A.~de Deus.
\newblock The construction of a boolean competitive neural network using ideas
  from immunology.
\newblock {\em Neurocomputing}, 50:51--85, 2003.

\bibitem{descioli-kurzban-todd-15IX}
Peter DeScioli, Robert Kurzban, and Peter~M. Todd.
\newblock Evolved decision makers in organizations.
\newblock In Stephen~M. Colarelli and Richard~D. Harvey, editors, {\em The
  Biological Foundations of Organizational Behavior}, chapter~IX, pages
  203--235. University of Chicago Press, Chicago, 2015.

\bibitem{edmonds-08VIII}
Bruce Edmonds.
\newblock A brief survey of some results on mechanisms and emergent outcomes.
\newblock In {\em Agent Cognitive Ability and Orders of Emergence: AISB 2008
  Proceedings, Volume 6}, pages 38--42, Aberdeen, 2008. The Society for the
  Study of Artificial Intelligence and Simulation of Behaviour.

\bibitem{edmonds-moss-05XI}
Bruce Edmonds and Scott Moss.
\newblock From kiss to kids: An `anti-simplistic' modelling approach.
\newblock In Paul Davidsson, Brian Logan, and Keiki Takadama, editors, {\em
  Multi-Agent and Multi-Agent-Based Simulation}, chapter~XI, pages 130--144.
  Springer Verlag, Berlin-Heidelberg, 2005.

\bibitem{epstein-axtell-96}
Joshua~M. Epstein and Robert Axtell.
\newblock {\em Growing Artificial Societies: Social science from the bottom
  up}.
\newblock Brookings Institution Press and The MIT Press, Washington D.C. and
  Cambridge (MA), 1996.

\bibitem{farmer-90PD}
J.~Doyne Farmer.
\newblock A rosetta stone for connectionism.
\newblock {\em Physica D}, 42(1-3):153--187, 1990.

\bibitem{farmer-patelli-zovko-05PNAS}
J.~Doyne Farmer, Paolo Patelli, and Ilija~I. Zovko.
\newblock Team assembly mechanisms determine collaboration network structure
  and team performance.
\newblock {\em Science}, 102(6):2254--2259, 2005.

\bibitem{fiol-lyles-85AMR}
C.~Marlene Fiol and Marjorie~A. Lyles.
\newblock Organizational learning.
\newblock {\em Academy of Management Review}, 10(4):803--813, 1985.

\bibitem{fioretti-12ORM}
Guido Fioretti.
\newblock Agent-based simulation models in organization science.
\newblock {\em Organizational Research Methods}, 16(2):227--242, 2012.

\bibitem{fioretti-16II}
Guido Fioretti.
\newblock Emergent organizations.
\newblock In Davide Secchi and Martin Neumann, editors, {\em Agent-Based
  Simulation of Organizational Behavior}, chapter~II, pages 19--41. Springer
  Verlag, Cham Heidelberg, 2016.

\bibitem{flyvbjerg-24PMJ}
Bent Flyvbjerg.
\newblock Heuristics for better project leadership: Teasing out tacit
  knowledge.
\newblock {\em Project Management Journal}, 55(6):615--625, 2024.

\bibitem{forbes-igboekwu-mousavi-20}
William~P. Forbes, Aloysius~Obinna Igboekwu, and Shabnam Mousavi.
\newblock {\em A Fast and Frugal Finance}.
\newblock Academic Press, London, 2020.

\bibitem{fort-24I}
Hugo Fort.
\newblock {\em From Growth Equations for a Single Species to Lotka-Volterra
  Equations for Two Interacting Species}, chapter~I, pages 1--1--1--47.
\newblock IOP Publishing, Bristol, 2024.

\bibitem{gabisch-lorenz-89}
G\"unter Gabisch and Hans-Walter Lorenz.
\newblock {\em Business Cycle Theory: A survey of methods and concepts}.
\newblock Springer Verlag, Berlin-Heidelberg, 1989.

\bibitem{garciadiaz-24NDPLS}
C\'esar Garc\'{i}a-D\'{i}az.
\newblock Resilience as anticipation in organizational systems: An agent-based
  computational approach.
\newblock {\em Nonlinear Dynamics, Psychology, and Life Sciences},
  28(3):409--429, 2024.

\bibitem{gigerenzer-02XXII}
Gerd Gigerenzer.
\newblock The adaptive toolbox: Toward a darwinian rationality.
\newblock In Lars Baeckman and Claes von Hofsten, editors, {\em Psychology at
  the Turn of the Millennium}, volume~1, chapter XXII, pages 458--482.
  Psychology Press, Hove and New York, 2002.

\bibitem{gigerenzer-16I}
Gerd Gigerenzer.
\newblock Rationality without optimization: Bounded rationality.
\newblock In Laura Macchi, Maria Bagassi, and Riccardo Viale, editors, {\em
  Cognitive Unconscious and Human Rationality}, chapter~I, pages 3--22. The MIT
  Press, Cambridge (MA), 2016.

\bibitem{gigerenzer-todd-99}
Gerd Gigerenzer, Peter~M. Todd, and The ABC~Research Group.
\newblock {\em Simple Heuristics That Make Us Smart}.
\newblock Oxford University Press, Oxford, 1999.

\bibitem{gjerstad-shachat-21NDPLS}
Steven Gjerstad and Jason~M. Shachat.
\newblock Individual rationality and market efficiency.
\newblock {\em Nonlinear Dynamics, Psychology, and Life Sciences},
  25(4):395--406, 2021.

\bibitem{gode-sunder-93JPE}
Dhananjay~K. Gode and Shyam Sunder.
\newblock Allocative efficiency of markets with zero-intelligence traders:
  Market as a partial substitute for individual rationality.
\newblock {\em Journal of Political Economy}, 101(1):119--137, 1993.

\bibitem{gode-sunder-97QJE}
Dhananjay~K. Gode and Shyam Sunder.
\newblock What makes markets allocationally efficient?
\newblock {\em The Quarterly Journal of Economics}, 112(2):603--630, 1997.

\bibitem{goodwin-67IV}
Richard~M. Goodwin.
\newblock A growth cycle.
\newblock In Charles~H. Feinstein, editor, {\em Socialism, Capitalism and
  Economic Growth}, chapter~4, pages 54--58. Cambridge University Press,
  Cambridge, 1967.

\bibitem{gordon-89AER}
Robert~J. Gordon.
\newblock Hysteresis in history: Was there ever a phillips curve?
\newblock {\em The American Economic Review}, 79(2):220--225, 1989.

\bibitem{gou-06JASSS}
Chengling Gou.
\newblock The simulation of financial markets by an agent-based mix-game model.
\newblock {\em Journal of Artificial Societies and Social Simulation}, 9(3),
  2006.

\bibitem{wade-hands-14JEM}
D.~Wade Hands.
\newblock Normative ecological rationality: Normative rationality in the
  fast-and-frugal heuristics research program.
\newblock {\em Journal of Economic Methodology}, 21(4):396--410, 2014.

\bibitem{herrmann-demiralp-05CN}
C.S. Herrmann and T.~Demiralp.
\newblock Human eeg gamma oscillations in neuropsychiatric disorders.
\newblock {\em Clinical Neurophysiology}, 116(12):2719--2733, 2005.

\bibitem{hertwig-herzog-09SC}
Ralph Hertwig and Stefan~M. Herzog.
\newblock Fast and frugal heuristics: Tools of social rationality.
\newblock {\em Social Cognition}, 27(5):661--698, 2009.

\bibitem{hidayati-kurniawan-21IJERSS}
Tri Hidayati and Wiwit Kurniawan.
\newblock Stability analysis of lotka-volterra model in the case of interaction
  of local religion and official religion.
\newblock {\em International Journal of Educational Research \& Social
  Sciences}, 2(3):542--546, 2021.

\bibitem{hong-page-09JET}
Lu~Hong and Scott Page.
\newblock Interpreted and generated signals.
\newblock {\em Journal of Economic Theory}, 144(5):2174--2196, 2009.

\bibitem{hong-page-12III}
Lu~Hong and Scott Page.
\newblock Some microfoundations of collective wisdom.
\newblock In H\'el\`ene Landemore and Jon Elster, editors, {\em Collective
  Wisdom: Principles and mechanisms}, chapter III, pages 56--71. Cambridge
  University Press, Cambridge, 2012.

\bibitem{hutchinson-gigerenzer-05BP}
John~M.C. Hutchinson and Gerd Gigerenzer.
\newblock Simple heuristics and rules of thumb: Where psychologists and
  behavioural biologists might meet.
\newblock {\em Behavioural Processes}, 69(2):97--124, 2005.

\bibitem{johnson-00ANYAS}
Norman~L. Johnson.
\newblock Importance of diversity: Reconciling natural selection and
  noncompetitive processes.
\newblock {\em Annals of the New York Academy of Sciences}, 901(1):54--66,
  2000.

\bibitem{kephart-hogg-huberman-90PD}
Jeffrey~O. Kephart, Tad Hogg, and Bernardo~A. Huberman.
\newblock Collective behavior of predictive agents.
\newblock {\em Physica D}, 42(1-3):48--65, 1990.

\bibitem{kephart-huberman-hogg-92IV}
Jeffrey~O. Kephart, Bernardo~A. Huberman, and Tad Hogg.
\newblock Can predictive agents prevent chaos?
\newblock In Paul Bourgine and Bernard Walliser, editors, {\em Economics and
  Cognitive Science}, chapter~IV, pages 41--55. Pergamon Press, Oxford, 1992.

\bibitem{klein-98}
Gary~A. Klein.
\newblock {\em Sources of Power: How people make decisions}.
\newblock The MIT Press, Cambridge (MA), 1998.

\bibitem{kohn-99}
Alfie Kohn.
\newblock {\em Punished by rewards: The trouble with gold stars, incentive
  plans, A's, praise, and other bribes}.
\newblock Houghton-Mifflin, Boston, 1999.

\bibitem{li-yang-shi-21H}
Chao-Hua Li, Wen-Goang Yang, and I-Tung Shih.
\newblock Exploration on the gap of single- and double-loop learning of
  balanced scorecard and organizational performance in a health organization.
\newblock {\em Heliyon}, 7(e08553):1--10, 2021.

\bibitem{liu-fan-17JNS}
Meng Liu and Meng Fan.
\newblock Permanence of stochastic lotka-volterra systems.
\newblock {\em Journal of Nonlinear Science}, 27(2):425--452, 2017.

\bibitem{lorenz-72AAAS}
Edward~N. Lorenz.
\newblock Predictability: Does the flap of a butterfly's wings in brazil set
  off a tornado in texas?
\newblock In {\em American Association for the Advancement of Science, 139th
  meeting}, Massachusets Institute of Technology, 1972.

\bibitem{lotka-25}
Alfred~J. Lotka.
\newblock {\em Elements of Physical Biology}.
\newblock Williams \& Wilkins Company, Baltimore, 1925.

\bibitem{lucas-76}
Robert~E. Lucas.
\newblock Econometric policy evaluation: A critique.
\newblock In Karl Brunner and Alan Meltzer, editors, {\em The Phillips Curve
  and Labor Markets}, pages 19--46. American Elsevier, New York, 1976.

\bibitem{malcai-biham-richmond-solomon-02PRE}
Ofer Malcai, Ofer Biham, Peter Richmond, and Sorin Solomon.
\newblock Theoretical analysis and simulations of the generalized
  lotka-volterra model.
\newblock {\em Physical Review E}, 66(3):031102, 2002.

\bibitem{mao-dworkin-suri-watts-17NC}
Andrew Mao, Lili Dworkin, Siddharth Suri, and Duncan~J. Watts.
\newblock Resilient cooperators stabilize long-run cooperation in the finetly
  repeated prisoner's dilemma.
\newblock {\em Nature Communications}, 8:13800, 2017.

\bibitem{mao-sabanis-renshaw-03JMAA}
Xuerong Mao, Sotirios Sabanis, and Eric Renshaw.
\newblock Asymptotic behaviour of the stochastic lotka-volterra model.
\newblock {\em Journal of Mathematical Analysis and Applications},
  287(1):141--156, 2003.

\bibitem{marasco-picucci-romano-16TFSC}
Addolorata Marasco, Antonella Picucci, and Alessandro Romano.
\newblock Market share dynamics using lotka-volterra models.
\newblock {\em Technological Forecasting and Social Change}, 105:49--62, 2016.

\bibitem{marasco-romano-18QQ}
Addolorata Marasco and Alessandro Romano.
\newblock Deterministic modeling in scenario forecasting: estimating the
  effects of two public policies on intergenerational conflict.
\newblock {\em Quality and Quantity}, 52(5):2345--2371, 2018.

\bibitem{marx-67}
Karl Marx.
\newblock {\em Das Kapital: Kritik del politischen Oekonomie}.
\newblock Otto Meissner, Hamburg, 1867.

\bibitem{miner-mezias-96OS}
Anne~S. Miner and Stephen~J. Mezias.
\newblock Ugly duckling no more: Past and futures of organizational learning
  research.
\newblock {\em Organization Science}, 7(1):88--89, 1996.

\bibitem{morris-pratt-03TFSC}
Steven~A. Morris and David Pratt.
\newblock Analysis of the lotka-volterra competition equations as a
  technological substitution model.
\newblock {\em Technological Forecasting and Social Change}, 70(2):103--133,
  2003.

\bibitem{mousavi-gigerenzer-17HO}
Shabnam Mousavi and Gerd Gigerenzer.
\newblock Heuristics are tools for uncertainty.
\newblock {\em Homo Oeconomicus}, 34(4):361--379, 2017.

\bibitem{mousavi-schulkin-19XIII}
Shabnan Mousavi and Jay Schulkin.
\newblock Ecological rationality and evolutionary medicine: A bridge to medical
  education.
\newblock In Jay Schulkin and Michael Power, editors, {\em Integrating
  Evolutionary Biology into Medical Education}, chapter XIII, pages 232--248.
  Oxford University Press, Oxford, 2019.

\bibitem{ormerod-08V}
Paul Ormerod.
\newblock What can agents know? the feasibility of advanced cognition in social
  and economic systems.
\newblock In Frank Guerin and Wamberto W M P~D Vasconcelos, editors, {\em Agent
  Cognitive Ability and Orders of Emergence (AISB 2008 Proceedings Volume 6)},
  pages 17--20, Aberdeen, 2008.

\bibitem{ortenblad-koris-13IJEM}
Anders \"Ortenblad and Riina Koris.
\newblock Is the learning organization idea relevant to higher education
  institutions? a literature review and a ``multi-stakeholder contingency
  approach''.
\newblock {\em International Journal of Educational Management},
  28(2):173--214, 2013.

\bibitem{othman-08CVI}
Abraham Othman.
\newblock Zero-intelligence agents in prediction markets.
\newblock In Lin Padgham, David Parkes, J\"org M\"uller, and Simon Parsons,
  editors, {\em Proceedings of the 7th International Conference on Autonomous
  Agents and Multiagent Systems (AAMAS 2008)}, pages 879--886, Estoril, 2008.

\bibitem{vanoverwalle-11XVI}
Frank~Van Overwalle.
\newblock Social learning and connectionism.
\newblock In Todd~R. Schachtman and Steve~S. Reilly, editors, {\em Associative
  Learning and Conditioning Theory: Human and non-human applications}, chapter
  XVI, pages 345--375. Oxford University Press, New York, 2011.

\bibitem{parker-kamenev-09PRE}
Matthew Parker and Alex Kamenev.
\newblock Extinction in the lotka-volterra model.
\newblock {\em Physical Review E}, 80:021129, 2009.

\bibitem{raab-gigerenzer-15FP}
Markus Raab and Gerd Gigerenzer.
\newblock The power of simplicity: A fast-and-frugal heuristics approach to
  performance science.
\newblock {\em Frontiers in Psychology}, 6(1672), 2015.

\bibitem{read-miller-98tutto}
Stephen~J. Read and Lynn~C. Miller, editors.
\newblock {\em Connectionist Models of Social Reasoning and Social Behavior}.
\newblock Lawrence Erlbaum Associates, M, 1998.

\bibitem{reichenbach-mobilia-frey-06PRE}
Tobias Reichenbach, Mauro Mobilia, and Erwin Frey.
\newblock Coexistence versus extinction in the stochastic cyclic lotka-volterra
  model.
\newblock {\em Physical Review E}, 74:051907, 2006.

\bibitem{rieskamp-hertwig-todd-06XI}
J\"org Rieskamp, Ralph Hertwig, and Peter~M. Todd.
\newblock Bounded rationality: Two interpretations from psychology.
\newblock In Morris Altman, editor, {\em Handbook of Contemporary Behavioral
  Economics: Foundations and Developments}, chapter~XI, pages 218--236. Sharpe,
  New York, 2006.

\bibitem{saffre-hildmann-deneubourg-18SI}
Fabrice Saffre, Hanno Hildmann, and Jean-Louis Deneubourg.
\newblock Can individual heterogeneity influence self-organized patterns in the
  termite nest construction model?
\newblock {\em Swarm Intelligence}, 12(2):101--110, 2018.

\bibitem{samuelson-39RES}
Paul~A. Samuelson.
\newblock Interactions between the multiplier analysis and the principle of
  acceleration.
\newblock {\em The Review of Economic Statistics}, 21(2):75--78, 1939.

\bibitem{sanchezpuig-zapata-pineda-iniguez-gershenson-23FCS}
Fernanda S\'anchez-Puig, Octavio Zapata, Omar~K. Pineda, Gerardo~I\ niguez, and
  Carlos Gershenson.
\newblock Heterogeneity extends criticality.
\newblock {\em Frontiers in Complex Systems}, 1:1111486, 2023.

\bibitem{scheutz-schermerhorn-03LIV}
Matthias Scheutz and Paul Schermerhorn.
\newblock Many is more, but not too many: Dimensions of cooperation of agents
  with and without predictive capabilities.
\newblock In Jiming Liu, Boi Faltings, Ning Zhong, Ruqian Lu, and Toyoaki
  Nishida, editors, {\em IEEE/WIC International Conference on Intelligent Agent
  Technology (IAT 2003)}, pages 378--384, Halifax, 2003.

\bibitem{senge-90}
Peter~M. Senge.
\newblock {\em The Fifth Discipline: The art \& practice of the learning
  organization}.
\newblock Doubleday, New York, 1990.

\bibitem{simon-57}
Herbert~A. Simon.
\newblock {\em Administrative Behavior}.
\newblock McMillan, New York, 1957.

\bibitem{simon-82}
Herbert~A. Simon.
\newblock {\em Models of Bounded Rationality}.
\newblock The MIT Press, Cambridge (MA), 1982.

\bibitem{skovranek-19M}
Tomas Skovranek.
\newblock The mittag-leffler fitting of the phillips curve.
\newblock {\em Mathematics}, 7(7):589, 2019.

\bibitem{smolensky-88BBS}
Paul Smolensky.
\newblock On the proper treatment of connectionism.
\newblock {\em Behavioral and Brain Sciences}, 11(1):1--74, 1988.

\bibitem{solomon-richmond-02EPJB}
Sorin Solomon and Peter Richmond.
\newblock Stable power laws in variable economies; lotka-volterra implies
  pareto-zipf.
\newblock {\em The European Physical Journal B}, 27(2):257--261, 2002.

\bibitem{sun-lorscheid-millington-lauf-magliocca-groeneveld-balbi-nolzen-muller-schulze-buchmann-16EMS}
Zhanli Sun, Iris Lorscheid, James D.~Millington andSteffen Lauf, Nicholas~R.
  Magliocca, J\"urgen Groeneveld, Stefano Balbi, Henning Nolzen, Birgit
  M\"uller, Jule Schulze, and Carsten~M. Buchmann.
\newblock Simple or complicated agent-based models? a complicated issue.
\newblock {\em Environmental Modelling \& Software}, 86:56--67, 2016.

\bibitem{takeuchi-96}
Yasuhiro Takeuchi.
\newblock {\em Global Dynamical Properties of Lotka-Voltera Systems}.
\newblock World Scientific, Singapore, 1996.

\bibitem{tchelebi-17OSD}
Nadine Tchelebi.
\newblock Taking bion ``back to basics'': Let us stop counting --- ``oneness''
  as the only basic-assumption mentality.
\newblock {\em Organizational \& Social Dynamics}, 17(1):50--70, 2017.

\bibitem{terna-02VIII}
Pietro Terna.
\newblock Cognitive agents behaving in a simple stock market structure.
\newblock In Francesco Luna and Alessandro Perrone, editors, {\em Agent-Based
  Methods in Economics and Finance: Simulations in Swarm}, chapter VIII, pages
  187--227. Kluwer Academic Publishers, Boston, 2002.

\bibitem{thierry-sheeren-marilleau-corson-almaric-monteil-15EM}
Hugo Thierry, David Sheeren, Nicolas Marilleau, Nathalie Corson, Marion
  Almaric, and Claude Monteil.
\newblock From the lotka-volterra model to a spatialized population-driven
  individual-based model.
\newblock {\em Ecological Modelling}, 306:287--293, 2015.

\bibitem{trianni-tuci-passino-marshall-11SI}
Vito Trianni, Elio Tuci, and Kevin~M. Passino.
\newblock Swarm cognition: An interdisciplinary approach to the study of
  self-organising biological collectives.
\newblock {\em Swarm Intelligence}, 5(1):3--18, 2011.

\bibitem{turner-75E}
James~T. Turner.
\newblock The stability and the intrinsic growth rates of prey and predator
  populations.
\newblock {\em Ecology}, 56(4):855--867, 1975.

\bibitem{vadillo-19AMC}
Fernando Vadillo.
\newblock Comparing stochastic lotka-volterra predator-prey models.
\newblock {\em Applied Mathematics and Computation}, 360:181--189, 2019.

\bibitem{visser-03JHBS}
Max Visser.
\newblock Gregory bateson on deutero-learning and double bind: A brief
  conceptual history.
\newblock {\em Journal of History of the Behavioral Sciences}, 39(3):269--278,
  2003.

\bibitem{visser-04WP}
Max Visser.
\newblock Deutero-learning in organizations: A review and a reformulation.
\newblock Technical Report RRM-2004-07-MGT, Radboud University Nijmegen, 2004.

\bibitem{visser-07AMR}
Max Visser.
\newblock Deutero-learning in organizations: A review and a reformulation.
\newblock {\em The Academy of Management Review}, 32(2):659--667, 2007.

\bibitem{vitanov-dimitrova-ausloos-10PA}
Nikolay~K. Vitanov, Zlatinka~I. Dimitrova, and Marcel Ausloos.
\newblock Verhulst-lotka-volterra (vlv) model of ideological struggle.
\newblock {\em Physica A}, 389(21):4970--4980, 2010.

\bibitem{volterra-26N}
Vito Volterra.
\newblock Fluctuations in the abundance of a species considered mathematically.
\newblock {\em Nature}, 118(2972):558--560, 1926.

\bibitem{walker-18}
Sam Walker.
\newblock {\em The Captain Class}.
\newblock Ebury Press, London, 2018.

\bibitem{wang-chen-wu-21DDNS}
Sheng-Yuan Wang, Wan-Ming Chen, and Xiao-Lan Wu.
\newblock Competition analysis on industry populations based on a
  three-dimensional lotka-volterra model.
\newblock {\em Discrete Dynamics in Nature and Society}, page 9935127, 2021.

\bibitem{weick-roberts-93ASQ}
Karl~E. Weick and Karlene~H. Roberts.
\newblock Collective mind in organizations: Heedful interrelating on flight
  decks.
\newblock {\em Administrative Science Quarterly}, 38(3):357--381, 1993.

\bibitem{wijeratne-yi-wei-09CSF}
A.W. Wijeratne, Fengqi Yi, and Junjie Wei.
\newblock Bifurcation analysis in the diffusive lotka–volterra system: An
  application to market economy.
\newblock {\em Chaos, Solitons \& Fractals}, 40(2):902--911, 2009.

\bibitem{wilensky-97}
Uri Wilensky.
\newblock Wolf sheep predation.
\newblock Technical report, Northwestern University, Center for Connected
  Learning and Computer-Based Modeling, Evanston, 1997.
\newblock http://ccl.northwestern.edu/netlogo/models/WolfSheepPredation.

\bibitem{wilensky-reisman-06CI}
Uri Wilensky and Kenneth Reisman.
\newblock Thinking like a wolf, a sheep, or a firefly: Learning biology through
  constructiing and testing computational theories -- an embodied modeling
  approach.
\newblock {\em Cognition and Instruction}, 24(2):171--209, 2006.

\bibitem{wilson-96TPB}
William~G. Wilson.
\newblock Lotka's game in predator-prey theory: Linking populations to
  individuals.
\newblock {\em Theoretical Population Biology}, 50(3):368--393, 1996.

\bibitem{wilson-98AN}
William~G. Wilson.
\newblock Resolving discrepancies between deterministic population models and
  individual-based simulations.
\newblock {\em The American Naturalist}, 151(2):116--134, 1998.

\bibitem{wong-wei-yang-tjosvold-17JWB}
Alfred Wong, Lu~Wei, Jie Yang, and Dean Tjosvold.
\newblock Productivity and participation values for cooperative goals to limit
  free riding and promote performance in international joint ventures.
\newblock {\em Journal of World Business}, 52(6):819--830, 2017.

\bibitem{williamswooley-chabris-pentland-hashmi-malone-10S}
Anita~Williams Wooley, Christopher Chabris, Alex Pentland, Nada Hashmi, and
  Thomas~W. Malone.
\newblock Evidence for a collective intelligence factor in the performance of
  human groups.
\newblock {\em Science}, 330(6004):686--688, 2010.

\bibitem{wu-liu-wang-12TFSC}
Lifeng Wu, Sifeng Liu, and Yinao Wang.
\newblock Grey lotka–volterra model and its application.
\newblock {\em Technological Forecasting and Social Change}, 79(9):1720--1730,
  2012.

\bibitem{yan-chu-06JCAM}
Xiang-Ping Yan and Yan-Dong Chu.
\newblock Stability and bifurcation analysis for a delayed lotka-volterra
  predator-prey system.
\newblock {\em Journal of Computational and Applied Mathematicals},
  196(1):198--210, 2006.

\bibitem{zhen-ma-02NA}
Jin Zhen and Zhien Ma.
\newblock Stability for a competitive lotka-volterra system with delays.
\newblock {\em Nonlinear Analysis}, 51(7):1131--1142, 2002.

\bibitem{zhu-yin-09JMAA}
Chao Zhu and George Yin.
\newblock On competitive lotka-volterra model in random environments.
\newblock {\em Journal of Mathematical Analysis and Applications},
  357(1):154--170, 2009.

\end{thebibliography}
\end{document}